%% file: NMRsuperconductorReview_epsv1.tex
\documentclass[12pt]{iopart}
\usepackage{amssymb}
\usepackage{graphicx}

\newcommand{\slrr}      {$T_1^{-1}$}

\newcommand{\cecoin}    {CeCoIn$_5$}

\newcommand{\cecoincdx}   {CeCo(In$_{1-x}$Cd$_{x})_5$}

\newcommand{\cecoincdfifteen}   {CeCo(In$_{0.85}$Cd$_{0.15}$)$_5$}
\newcommand{\cerhin}    {CeRhIn$_5$}

\newcommand{\tc}     {$T_{\rm c}$}
\newcommand{\slrrtext}  {spin lattice relaxation rate}

\newcommand{\hightc}   {high-T$_{\rm c}$}

\newcommand{\cecusi}    {CeCu$_2$Si$_2$}
\newcommand{\pucoga}    {PuCoGa$_5$}
\newcommand{\purhga}    {PuRhGa$_5$}

\begin{document}

\title[NMR in heavy fermion superconductors]{Nuclear magnetic resonance in the heavy fermion superconductors}

\author{N. J. Curro}

\address{Department of Physics,
University of California, One Shields Avenue, Davis, CA 95616-8677}
\ead{curro@physics.ucdavis.edu}

\begin{abstract}
Nuclear magnetic resonance has emerged as a vital technique for investigating strongly correlated electron systems, and is particularly important for studying superconductivity.  In this paper the basic features of NMR as a technique for probing the superconducting state are reviewed.  Topics include include spin relaxation processes,
studies of vortex lattices, and phenomena associated with unconventional pairing symmetries.  Recent experimental work is reviewed, with a particular emphasis on the heavy fermion superconductors.
\end{abstract}

\submitto{\RPP}
\maketitle

\tableofcontents
\newpage
\title[NMR in heavy fermion superconductors]{Nuclear magnetic resonance in the heavy fermion superconductors}
\section{Introduction}

Nuclear Magnetic Resonance (NMR) has impacted several branches of science, not
the least of which is the study of strongly correlated electrons in
condensed matter. The pioneering works of Hebel and Slichter
\cite{hebelslichter1, hebelslichter2} and Masuda and Redfield
\cite{masudaredfield} revealed the power of this technique to the
condensed matter community. These NMR experiments measuring the spin
lattice relaxation rate in the superconducting state of Al showed
the presence of a superconducting gap, and more importantly
that the Cooper pairs are a coherent superposition of two particles.
Their results clearly demonstrated that the two-fluid picture of
superconductivity, consisting of single-particle states, was
untenable. Indeed, in their seminal paper on the theory of superconductivity, Bardeen Cooper and Schrieffer point out the role of both NMR experiments and ultrasonic attenuation in support of their theory \cite{bardeencooperschrieffer}.

NMR played a central role in the study of the \hightc\ cuprates, and continues to yield important information about novel superconductors \cite{CPSreview,GrafePRLLaOFFeAs}.  In the last decade or so, there has been a renaissance in the study of the heavy fermion superconductors.  These materials were first discovered in 1979, but were largely overshadowed by the tremendous effort focused on \hightc\ materials \cite{steglichdiscovery,zachreview,kitaokareview}.  Recent discoveries of the Ce$_{n}$M$_m$In$_{3n+2m}$ family of compounds, various actinide-based superconductors, the breakdown of Fermi-liquid theory, and the emergence of superconductivity in the vicinity of quantum phase transitions have turned the attention of the condensed matter community to the nature of the superconductivity, magnetism, and new broken symmetries that emerge in these compounds.  This article is intended to briefly review some of the important contributions of NMR to the study of these phenomena, with an emphasis on a discussion of NMR as a general technique for investigating superconductivity rather than a comprehensive review of the literature.  In particular, this article is directed at non-NMR specialists who wish to understand and critically evaluate new data.  For an in-depth review of NMR in heavy fermions, the reader is referred to Ref. \cite{KitaokaBook}.

Upon reflection, one might expect \textit{a priori} that the
Meissner effect in a superconductor should preclude any NMR
experiments.  Indeed, many NMR experiments in superconductors are
dominated by the response of the superconductor to magnetic fields.
NMR studies of superconductors can be divided into two categories:
those which investigate processes driven by (i) hyperfine fields and
by (ii) orbital supercurrents. Hyperfine fields arise from interactions
between the nuclear and electron spins.  In metals and
superconductors, the dominant contribution to relaxation of the
nuclear spins usually arises from scattering from the
quasiparticles.  Orbital (super)currents arise in superconductors as
a response to screen external magnetic fields. Typically these
currents are manifest in a vortex lattice, and give rise to an
inhomogeneous magnetic induction. Since the nuclear spins couple to
this magnetic induction through the Zeeman interaction, the NMR
spectrum is a sensitive probe of the physics of the vortex lattice.

In order to fully understand the behavior of nuclei in a
superconductor, it is crucial to first understand their behavior in
conventional metals.  The principles of NMR in metallic systems is discussed in sections \S\ref{sec:NMRbasics} and
\S\ref{sec:NMRmetals}. Section
\S\ref{sec:NMRSChyperfine} addresses the hyperfine
interaction in superconductors, which probes the low energy
properties of the superconducting gap function.  Section
\S\ref{sec:NMRSCorbital} focuses on measurements that couple to the
orbital supercurrents, and reveals physics of the vortex
lattice.  In section \S\ref{sec:NMRheavy} we review NMR measurements
in the superconducting state of the heavy fermion materials.  NMR
measurements in the high temperature superconductors has been
instrumental to elucidate the physics not only of the
superconducting state, but the low energy physics of the spin
system.  Aside for a brief discussion in \S\ref{sec:NMRSChyperfine}
on the measurement of the symmetry of the superconducting order
parameter in the cuprates via spin echo decay measurements, this
article will not attempt to review the vast literature of NMR in the
high temperature superconductors. We refer the reader to
\cite{CPSreview} for details.

\input{NMRbasics}

\input{NMRmetals}

\input{NMRSChyperfine}

\input{NMRSCheavyelectron}

\section*{Acknowledgements}
\addcontentsline{toc}{section}{Acknowledgements}

We thank D. Pines, L. Greene, R. R. Urbano and C. P. Slichter, for critical readings and enlightening discussions, and the Aspen Center for Physics for hospitality during the writing of a portion of this manuscript.

\newpage
\section*{References}
\addcontentsline{toc}{section}{References}
\bibliography{C:/Users/curro/Documents/Documents/Bibliography/CurroNMR}

\end{document}

%% file: NMRbasics.tex
\section{NMR Basics\label{sec:NMRbasics}}

\subsection{Hamiltonians and Spectra}

Nuclear spins in condensed matter constitute are a nearly ideal example of an isolated ensemble of coupled particles that experience a weak
coupling to an external bath. This weak coupling, the nuclear-electron coupling, gives rise to small perturbations of the nuclear spin Hamiltonian, as well as provides a channel for relaxation and decoherence of the nuclear spin system.  It is via this interaction that the nuclei can provide information about the electronic degrees of freedom in strongly correlated electron systems.  The nuclear spin Hamiltonian in correlated systems can be written as the sum of four interactions:
\begin{equation}
\mathcal{H}=\mathcal{H}_{\rm Z}+\mathcal{H}_{\rm Q} +
\mathcal{H}_{\rm dip} + \mathcal{H}_{\rm e-n}.
\label{eqn:hamiltonian}
\end{equation}
The first term describes the Zeeman interaction between the nuclear
spin $I$ and the local magnetic field, $\mathbf{H}_0$:
\begin{equation}
\mathcal{H}_{\rm Z} = \gamma\hbar \mathbf{H}_o\cdot\mathbf{\hat{I}},
\label{eqn:Hzeeman}
\end{equation}
where the gyromagnetic ratio $\gamma$ is related to the nuclear
moment $\mu$ by $\gamma\hbar = g\mu_N\sqrt{I(I+1))}$. Here $\mu_N$ is the
nuclear magneton, and $g$ is the nuclear g-factor. This term is
usually the dominant interaction for nuclei in an external field,
with $\gamma H_0\sim 100$ MHz. Since $h/k_B\approx 48 \mu{\rm K}/{\rm
MHz}$,  the nuclear spin system can
be treated in the high temperature limit, as most phenomena of
interest in correlated electrons are manifest at temperatures $> 1$ mK.

The second term in Eq. (\ref{eqn:hamiltonian}), the quadrupolar interaction or the charge part of the electron-nuclear interaction, is given by:
\begin{equation}
\mathcal{H}_{\rm Q} = \frac{eQV_{zz}}{4I(2I-1)}[(3\hat{I}_z^2 -
\hat{I}^2) + \eta(\hat{I}_x^2-\hat{I}_y^2)],
\label{eqn:Hquadrupolar}
\end{equation}
where $e$ is the electron charge, $Q$ is the quadrupolar moment,
$V_{\alpha\beta}$ are the components of the EFG tensor, and $\eta
=(V_{xx}-V_{yy})/V_{zz}$ is the asymmetry parameter of the EFG.  The
NQR frequency (in zero field) is defined as:  $\nu_Q =
\nu_z\sqrt{1+\eta^2/3}$, where $\nu_z=eQ/2I(2I-1)h$. For a site with
axial symmetry ($\eta=0$), there are $2I-1$ quadrupolar resonances
at frequencies $n\nu_Q$, where $n = 1, \dots ,2I-1$.  If $\eta>0$,
then the resonances are not equally spaced.  The EFG is fully
characterized by three parameters: $\nu_z, \eta$ and a unit vector,
$\hat{q}$, in the direction of the principle axis of the EFG with
the largest eigenvalue.  The quadrupolar interaction vanishes for
$I=1/2$, but can be significant for nuclei with sufficiently large
quadrupolar moments and $I> 1/2$.  Typically
$\langle\mathcal{H}_{\rm Q}\rangle/h\sim1-100$ MHz.  In practice,
this means that the degeneracy of the nuclear spin manifold is
lifted even in zero field, and one can detect resonances in the absence of an applied field. This technique is known as Nuclear Quadrupolar Resonance (NQR), whereas the term NMR is reserved for resonance experiments in field. This is particularly important for superconductors, since it
allows one to measure the spin lattice relaxation in zero field,
without the influence of the vortex lattice.

The third interaction in Eq. (\ref{eqn:hamiltonian}) is the dipolar interaction between nuclear spins in a lattice, and is given by:
\begin{equation}
\mathcal{H}_{\rm dip} = \frac{1}{2}\sum_{j,k=1}^N\left(\frac{\mathbf{\mu}_j\cdot\mathbf{\mu}_k}{r_{jk}^3}-\frac{3(\mathbf{\mu}_j\cdot\mathbf{r}_{jk})(\mathbf{\mu}_k\cdot\mathbf{r}_{jk})}{r_{jk}^5}\right),
\label{eqn:Hdipolar}
\end{equation}
where $\mathbf{r}_{jk}$ is the distance between the nuclei moments $j$ and $k$ with moments $\mathbf{\mu}=\gamma\hbar\mathbf{\hat{I}}$. In most solids, the dipolar interaction gives rise to a static broadening of the lineshape, and in most cases the lineshape can be well approximated by a Gaussian with second moment given by:
\begin{equation}
\langle\Delta\omega^2\rangle=\frac{3}{4}\gamma^4\hbar^2I(I+1)\frac{1}{N}\sum_{j,k}\frac{(1-3\cos^2\theta_{jk})^2}{r_{jk}^6}. \label{eqn:dipolarsecondmoment}
\end{equation}
Usually $\langle\Delta\omega^2\rangle$ is temperature independent, and depends on the structural details of the particular compound of interest. Typically $\langle\Delta\omega^2\rangle \sim 1-10$ G$^2$.   For studies of vortex physics in superconductors, it is preferable that $\langle\Delta\omega^2\rangle$ is as small as possible, since the inhomogeneous field distribution in the mixed state of a type II superconductor broadens the lineshape in a quantifiable fashion.  Since this vortex lattice lineshape broadening is typically only on the order of a few Gauss, one is often faced with the fact that the intrinsic dipolar broadening limits the spectral resolution (see \S\ref{sec:NMRSCorbital}.1).  Furthermore, the quadrupolar interaction (\ref{eqn:Hquadrupolar}) can also contribute to line broadening, particularly when there are lattice strains which can give rise to distributions of EFGs.

The fourth term in Eq. (\ref{eqn:hamiltonian}) is the magnetic part of the nuclear-electron interaction, and is given by \cite{abragambook}:

\begin{equation}
\mathcal{H}_{\rm e-n} = \gamma\hbar\mathbf{\hat{I}}\cdot\sum_i(2\mu_B)\left(
\frac{\mathbf{l}_i}{r_i^3}-\frac{\mathbf{S}_i}{r^3_i}+3\frac{\mathbf{r}_i(\mathbf{S}_i\cdot\mathbf{r}_i)}{r_i^5}+\frac{8}{3}\pi\mathbf{S}_i\delta(\mathbf{r}_i)\right)
\label{eqn:en}
\end{equation}
where $\mathbf{S}_i$ and $\mathbf{l}_i$ are the spin and angular momentum (with respect to the nucleus) of the i$^{th}$ electron.  The first term of Eq. (\ref{eqn:en}) is the orbital interaction, which gives rise to the chemical shift in materials where orbital angular momentum is unquenched, as well as in superconductors with supercurrent distributions. The remaining three terms are the dipolar and contact part of the hyperfine interaction, which is usually the dominant relaxation mechanism in metals and superconductors.  Both terms can be understood in terms of extra fields experienced by the nucleus, either by a hyperfine field, $\mathbf{H}_{hf}$, or by a magnetization $\mathbf{M}_{orb}$ created by a current distribution.  The former typically is the primary interaction between nuclei and quasiparticles in superconductors, and the latter couples the nuclei to the vortex lattice in superconductors. In most strongly correlated electron systems,
the hyperfine interaction is complicated by the presence of \textit{transferred} hyperfine interactions, given by a complex exchange process involving orbital overlaps between different ions in a material. In practice, the hyperfine interaction is approximated as:
\begin{equation}
\mathcal{H}_{hf} = \mathbf{\hat{I}}\cdot\mathbf{A}\cdot\mathbf{S}(\mathbf{r}=\mathbf{0}) + \sum_i B_i  \mathbf{\hat{I}}\cdot\mathbf{S}(\mathbf{r}_i)
\label{eq:hyptransferred}
\end{equation}
where $\mathbf{A}$ is an on-site hyperfine coupling (typically dominated by core-polarization effects), and $B_i$ are transferred hyperfine coupling to the neighboring spins, $\mathbf{S}(\mathbf{r}_i)$.  These couplings are usually taken as empirical values, since accurate calculations of these values involves detailed electronic structure calculations, which are not always available, especially in strongly correlated systems \cite{Meier}.  Eq. (\ref{eq:hyptransferred}) is the basis of the Mila-Rice-Shastry Hamiltonian in the cuprates \cite{MilaRiceHamiltonian,shastrycuprates}, and is a good approximation to the interactions in most heavy electron systems \cite{CurroKSA}.

\subsection{CW versus pulse techniques}

Equation (\ref{eqn:hamiltonian}) lifts the degeneracy between the nuclear $\hat{I}_z$ levels, and the spectral properties can be understood in terms of the energy level differences between the eigenvalues of $\mathcal{H}$.  A basic NMR spectrometer measures the energy absorbed by an ensemble of nuclear spins as a function of either fixed field while frequency is swept, or vice versa.  Such an experiment describes a basic continuous-wave (CW) spectrum measurement.  Over the last three decades pulsed techniques have been developed such that a nuclear spin system can be excited by a single short pulse (of duration $\approx 1-10$ $\mu$sec).  The spectrum is then given by the Fourier transform of the response to the pulsed excitation.  A \textit{spin echo} is the response of a nuclear spin system to a series of two pulses.  The Fourier transform of a spin echo is also identical to the spectrum, with the advantage that the response of the system is separated temporally from the excitation pulse, eliminating a source of noise.

\subsection{Density Matrices}

The behavior of the nuclear spin system is best described by an ensemble averaged density matrix, $\hat{\rho}(t)=\overline{|\psi(t)\rangle\langle\psi(t)|}$, where $|\psi(t)\rangle$ is the wavefunction of a single nuclear spin in the ensemble.  The quantities measured in an NMR experiment, $M_x(t)$ or $M_z(t)$, describing the transverse or longitudinal nuclear magnetization, can be written in terms of the density matrix via the expression $M_\alpha(t)=\overline{\langle M_\alpha\rangle}=\textrm{Tr}\{\hat{M}_\alpha\cdot\hat{\rho}(t)\}$.  Details of the spectrum and relaxation rates are therefore contained in $\hat{\rho}(t)$.  For a typical spin 1/2 nucleus, the density matrix is given by:
\begin{eqnarray}
\hat{\rho}(t) = \nonumber\\
\left(
                   \begin{array}{cc}
                     \rho_{11}(0)e^{-t/T_1}+\rho_{11}^{EQ}(1-e^{-t/T_1}) & \rho_{12}(0)e^{(+i\omega-1/T_2) t} \\
                     \rho_{21}(0)e^{(-i\omega-1/T_2) t} & \rho_{22}(0)e^{-t/T_1}+\rho_{22}^{EQ}(1-e^{-t/T_1}) \\
                   \end{array}
                 \right),
\label{eqn:densitymatrix}
\end{eqnarray}
where $T_1^{-1}$ and $T_2^{-1}$ are the spin-lattice relaxation rate and the spin decoherence rates, respectively, $\hat{\rho}^{EQ} = \exp(-\hat{\mathcal{H}}/kT)/Z$ is the density matrix in thermal equilibrium, and $Z$ is the partition function.  Clearly, the relaxation rates tend to bring the density matrix to its equilibrium, time-independent value.  Spin-lattice relaxation is responsible for the time decay of the diagonal terms of the density matrix, whereas spin-decoherence relaxation is responsible for the decay of the off-diagonal terms.  The spin-lattice relaxation rate provides the time-dependence of the $M_z$ component of the nuclear spin magnetization, whereas the decoherence rate $T_2^{-1}$ affects the transverse magnetization, $M_x(t)$.

\subsection{Relaxation Processes}

The spin lattice relaxation rate arises from fluctuations which couple nuclear spin levels which have a finite power spectral density at the nuclear Larmor frequency, $\omega_L = \gamma H_0$.  The most straightforward method to calculate \slrr\ is via first order time-dependent perturbation theory, which gives the Fermi golden rule:
\begin{equation}
P_{i\rightarrow f} = \frac{2\pi}{\hbar}|\langle i|\mathcal{H}_1|f\rangle|^2\delta(\omega - E_f+E_i)
\end{equation}
where $P_{i\rightarrow f}$ is the rate of transitions between states $|i\rangle$ and $|f\rangle$, and $\mathcal{H}_1$ is the (time dependent) Hamiltonian giving rise to the spin lattice relaxation.  In heavy fermion materials, typically  $\mathcal{H}_1$ is given by hyperfine interaction (Eq. \ref{eq:hyptransferred}), where the time dependence arises from the dynamical fluctuations of the spins $\mathbf{S}(\mathbf{r}_i,t)$, or from coupling to the orbital magnetization in a vortex lattice, which may fluctuate due to vortex motion.   The principle of detailed balance assures that an ensemble of nuclear spins will reach equilibrium through a series of transitions between levels as described above.  In equilibrium, the population of the nuclear spin levels can be described by a spin temperature, $T_s$, such that the population of each level, $P_i$, is described by the Boltzmann factor: $P_i = \exp(-E_i/k_bT_s)/Z$, where $E_i$ is the energy of the i$^{th}$ level, and $Z$ is the partition function.  One can show that the time dependence of the spin temperature $T_s \equiv 1/k_B\beta_s$ can be described as:
\begin{equation}
\frac{d\beta_s}{dt}= \frac{1}{T_1}(\beta_s(t)-\beta_{EQ})
\end{equation}
where $\beta_{EQ}$ is the equilibrium spin temperature, and \slrr\ is the spin lattice relaxation rate \cite{CPSbook}.

In metallic systems, the dominant coupling to the nuclei is the hyperfine interaction, and the spin lattice relaxation rate can be written in terms of the autocorrelation function of the dynamical hyperfine field, $H_{\rm hf}(t)$ as:
\begin{equation}
\label{eqn:t1generalbasic} T_{1}^{-1}=\frac{\gamma^2}{2}\int_0^{\infty}{\langle H_{\rm hf}(t)H_{\rm hf}(0)\rangle e^{i\omega_0 t}}dt,
\end{equation}
where $\omega_0$ is the Larmor frequency.  If we assume that $H_{\rm hf}(t)$ fluctuates randomly between $\pm H_{\rm hf}$ with autocorrelation function: $\langle H_{\rm hf}(t)H_{\rm hf}(0)\rangle = H_{\rm hf}^2 e^{-t/\tau}$, then we can write:
\begin{equation}
T_{1}^{-1} = \frac{\gamma^2 H_{\rm hf}^2\tau}{1+\omega_0^2\tau^2}.
\label{eqn:t1bpp}
\end{equation}
This is a reasonable description of many processes involving phase transitions, where the fluctuation time scale changes by several orders of magnitude through a phase transition. The limit $\omega_0\tau\ll 1$ corresponds to motionally narrowed limit, as described below. Note that in this case, \slrr\ is independent of frequency.

\subsubsection{Bloch Equations}

By employing Eq. \ref{eqn:densitymatrix}, one can derive expressions for the expectation value of the nuclear magnetization as a function of time.  The expressions are the famous Bloch equations, given by:
\begin{eqnarray}
\frac{dM_z}{dt} &=& \gamma(\mathbf{M}\times\mathbf{H})_z+\frac{M_0-M_z}{T_1}\\
\frac{dM_x}{dt} &=& \gamma(\mathbf{M}\times\mathbf{H})_x- \frac{M_x}{T_2}\\
\frac{dM_y}{dt} &=& \gamma(\mathbf{M}\times\mathbf{H})_y-\frac{M_y}{T_2}.
\end{eqnarray}
These equations not only make a connection to the classical behavior of a precessing magnetic moment in a magnetic field, but they reveal a more physical picture of the relaxation rates \slrr\ and $T_2^{-1}$.  Clearly, \slrr\ corresponds to a longitudinal relaxation of the magnetization along the $z$-axis, whereas $T_2$ is a transverse relaxation in the $xy$-plane. Quantum mechanically, $T_2$  corresponds to a decoherence time, in which the ensemble of precessing nuclear spins looses its coherence. Immediately after a pulse, one creates off-diagonal coherences in the density matrix. Over a time scale, $T_2$, the nuclear wavefunctions are no longer in phase, and the precessing nuclear spins are no longer in step \cite{schlosshauer}.

\subsection{Motional Narrowing}

\begin{figure}
\centering
\includegraphics[width=\textwidth]{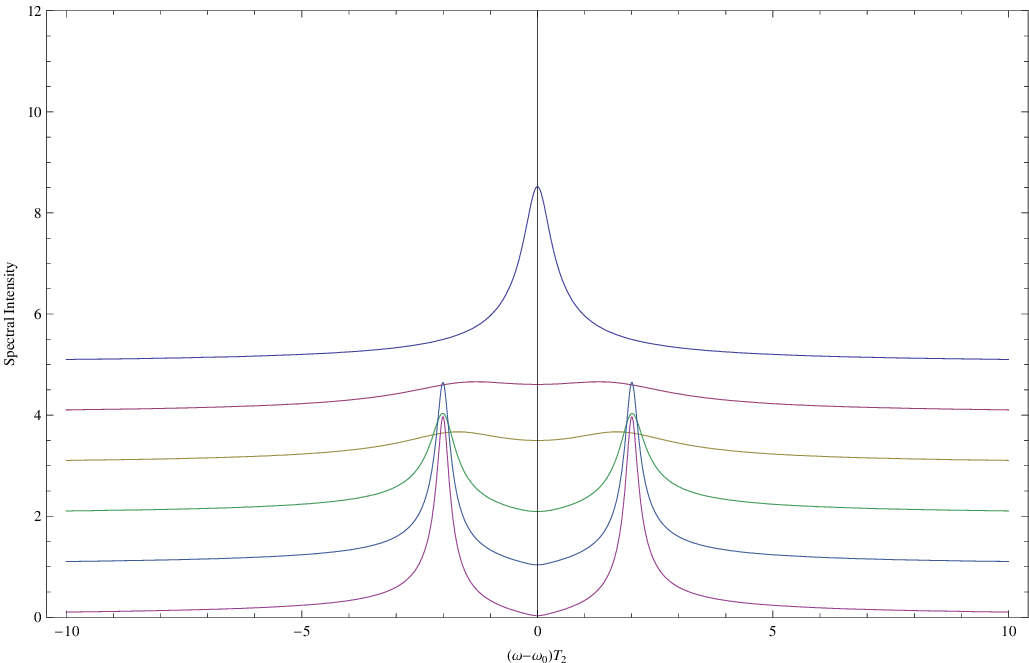}
\caption{Calculated spectra for a nucleus experiencing a dynamic hyperfine field $\pm h_{\rm hf}$ with correlation time $\tau$.  The graphs are offset vertically, with values of $\gamma h_{\rm hf}T_2=4$, and $T_2/\tau = 100,\,10,\, 8,\,1,\,0.1,\,0.01$ from the upper graph to the bottom graph. \label{fig:motionalnarrow} }
\end{figure}

When a nucleus experiences a static field, perhaps from an ordered magnetic phase, the NMR spectrum will reflect this internal field, usually as a broadening or a shift of the resonance frequency in the absence of this field.  On the other hand, if this field is fluctuating on a fast time scale, the effect will be averaged out and will not contribute to the NMR spectrum.  This phenomenon is known as motional narrowing, and plays a key role in the NMR physics of many heavy electron systems.  The relevant scale for the crossover from the static to the motionally narrowed limit is given by $\gamma h_{hf}\tau \sim 1$, where $h_{hf}$ is the magnitude of the fluctuating field, and $\tau$ is the correlation time of these fluctuations.  This phenomenon is shown in Fig. \ref{fig:motionalnarrow}, where the spectra are shown for a series of values of $\tau$, where the field hops randomly between $\pm h_{hf}$ with correlation time $\tau$.  A detailed description of the general phenomenon can be found in \cite{CPSbook}.

\begin{figure}
\centering
\includegraphics[width=\textwidth]{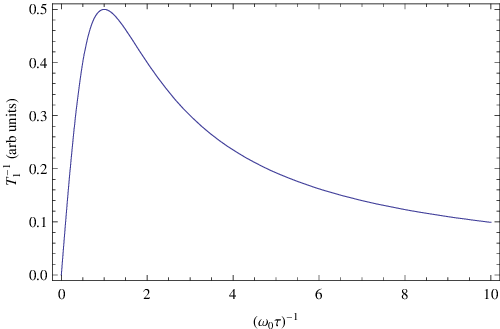}
\caption{The \slrrtext\ versus $(\omega_0\tau)^{-1}$ showing the Bloembergen-Purcell-Pound (BPP) peak (Eq. \ref{eqn:t1bpp}). \label{fig:t1bpp}}
\end{figure}

\subsection{NMR and magnetic phase transitions}

\begin{figure}
\centering
\includegraphics[width=\textwidth]{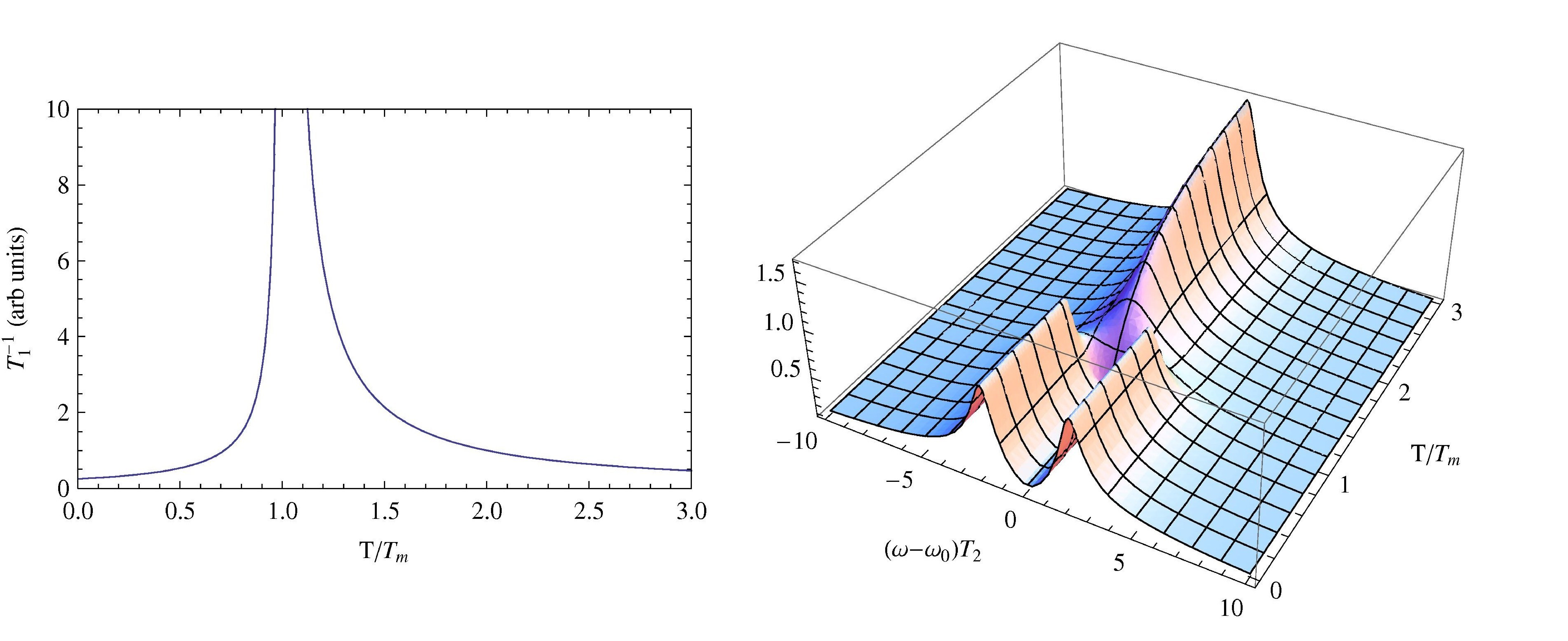}
\caption{Spectra and relaxation rate for an ensemble of nuclei experiencing a dynamic hyperfine field $\pm h_{\rm hf}$ as a function of temperature. Here we consider a correlation time $\tau=\tau_0(T/T_m-1)^{-\eta}$ for $T>T_m$ and $\eta=1.1$, where  $T_m$ represents a magnetic ordering temperature at which point the electron spins (and hence the hyperfine field at the nucleus) becomes static. Note that the temperature dependence of the spectra below $T_m$ differ from typical materials at a second order transition, where the ordered moment (and the hyperfine field) grow continuously from zero.  In this case, there is only a single value of $|h_{\rm hf}|$. \label{fig:t1andspectrum}}
\end{figure}

In a typical magnetic phase transition, both \slrr\ and the spectrum change dramatically. To illustrate this, let us assume that there is an internal magnetic field $h_{\rm hf}(t)$ from the magnetic order that is created by the hyperfine interaction at the nuclear site. Above the magnetic ordering temperature, $h_{\rm hf}(t)$ fluctuates randomly from $\pm h_0$ with an autocorrelation function $\langle h_{\rm hf}(t)h_{\rm hf}(0)\rangle = h_0^2e^{-t/\tau}$.  If the transition is second order, the temperature dependence of $\tau$ is typically described by $\tau(T)=\tau_1(1-T/T_m)^{-\eta_1}$ for $T<T_m$ and $\tau(T)=\tau_2(T/T_m-1)^{-\eta_2}$ for $T>T_m$.  When critical opalescence sets in above the phase transition, \slrr\ typically increases.  This can be seen in Fig. \ref{fig:t1bpp}, where \slrr\ is plotted as a function of $1/\tau$; i.e. high temperatures would correspond to large values of $1/\tau$.  As the correlation time grows and reaches a value on the order of the Larmor frequency, the fluctuating field becomes most efficient at inducing transitions between the nuclear levels and \slrr\ reaches a maximum.  Fig. \ref{fig:t1andspectrum} shows the spectrum and the spin lattice relaxation rate as a function of temperature assuming $\eta=1.1$.  Clearly, there are dramatic changes as the temperature goes through $T_m$, and static order sets in.  This scenario is a good model for a second order  antiferromagnetic phase transition, since the hyperfine field from the ordered spins is either positive or negative at the nuclear sites in the unit cell.  However, in real systems the ordered moment grows continuously from zero, whereas in the case studied here it is finite and temperature independent.

%% file: NMRmetals.tex
\section{NMR in Metals\label{sec:NMRmetals}}

\subsection{RF Radiation in Metals}
NMR in metallic systems is  experimentally challenging because the radiofrequency skin depth limits the effective volume of sample that can be measured.  This skin depth affects both the excitation pulses ($H_1$) and the signal from the precessing nuclei within the bulk.  The rf skin depth $\delta = \sqrt{2\rho/\omega\mu}$, where $\rho$ is the resistivity of the metal, is on the order of a few microns for good metals at typical NMR frequencies, whereas sample sizes are typically on the order of a few mm.  Nevertheless, the number of nuclei within the skin depth layer of a single crystal may reach the order of $10^{17}$, which is sufficient to detect.  Reducing the temperature and averaging several echoes (typically between 10 to 10$^5$) increase the signal-to-noise ratio (SNR) and make it possible to accurately measure spectra and relaxation rates.  Powdered polycrystalline samples increase the surface to volume ratio and boost the SNR significantly, but can be complicated by orientation-dependent terms such as the quadrupolar interaction for nuclei with spin greater than $I=\frac{1}{2}$.  To get around this, one can also work with aligned samples, in which the powder is mixed with epoxy and cured in a magnetic field.  If the sample has an anisotropic magnetic susceptibility, the crystallites will align in the external field.

\subsection{Hyperfine coupling and Knight shifts}

In a Fermi liquid the dominant coupling is the contact hyperfine coupling to the quasiparticle spins, given by:
\begin{eqnarray}
\mathcal{H}_{\rm hyp} &=& A\hat{\mathbf{I}}\cdot\mathbf{S}\nonumber\\
&=& A\hat{I}_zS_z+\frac{1}{2}A(\hat{I}_+S_-+\hat{I}_-S_+).
\end{eqnarray}
This coupling allows the nuclei probe both the static susceptibility, $\chi_0$, and the dynamical susceptibility, $\chi(\mathbf{q},\omega)$, of the quasiparticles.   The diagonal term $A\hat{I}_z\hat{S}_z$ gives rise to a static shift of the resonance frequency.  The Knight shift, $K$, which measures the percent shift of the resonance frequency from that of an isolated nucleus ($\omega_0=\gamma H_0$) is given by: $K = A\chi_0/\hbar\gamma\mu_B$.  In a Fermi liquid, $\chi_0$ is given by the Pauli susceptibility, so  $K \sim A N(0)$ is temperature independent.

This  scenario works well for simple metals such as Li and Na, yet there are many cases where the Knight shift is more complex.  In Pt, for example, there are multiple hyperfine couplings to the d- and sp- bands, and hence  several contributions to the total shift \cite{PtNMR}.  In many-electron atoms, there is also a core-polarization term, in which the core s electrons acquire a population difference between the up- and down-spin states.  This difference arises because the orthogonal eigenstates of the many-electron atom get mixed by the perturbing influence of the external field.  In practice, it is difficult to estimate the contribution of a core-polarization term versus a purely contact term \cite{walstedtYBCO}. As a result, hyperfine couplings are usually taken to be material dependent parameters.  In metals with local moments, such as rare-earth and d-electron systems, there is also a second hyperfine coupling to these moments that give rise to a strong temperature dependence of the total shift.

\subsection{Spin-flip scattering and spin-lattice relaxation}

The off diagonal terms of the hyperfine coupling can be written as: $\frac{A}{2}(\hat{I}_+\hat{S}_++\hat{I}_-\hat{S}_-)$.  This perturbation corresponds to a spin-flip exchange between the quasiparticle spin and the nuclear spin.   These processes do not shift the resonance frequency, but do affect the dynamics of the nuclei.  In fact, this process is crucial to bring about an equilibrium population distribution among the nuclear spin levels.  In a Fermi liquid, as the quasiparticles scatter from one nucleus to another, they maintain essentially the same energy since the nuclear Zeeman energy is orders of magnitude lower than the Fermi level, and the quasiparticle Zeeman energy can be absorbed by states within $k_B T$ of $E_F$. By using Fermi's Golden Rule one can show that the \slrrtext\ can be written as:
\begin{equation}
\label{eqn:t1general} T_{1}^{-1}=\gamma^2 A^2 \int_0^{\infty}{\langle \hat{S}_{+}(t)\hat{S}_{-}(0)\rangle e^{i\omega_0 t}}dt,
\end{equation}
where the brackets indicate a thermal averaged correlation function.
In a Fermi liquid, the states available for scattering the quasiparticles are those at the Fermi surface, and by a simple counting argument one can show that (\ref{eqn:t1general}) can be written as:
\begin{equation}
T_{1}^{-1} = \frac{\gamma}{2}\int_0^{\infty}N(E_i)N(E_f)f(E_i)(1-f(E_f))dE_i
\label{eqn:slrrmetal}
\end{equation}
where $E_f-E_i=\hbar\omega_0$.  Since $f(E)(1-f(E))\approx k_B T \delta(E-E_F)$, we find that $T_1^{-1}\sim T A^2 N^2(0)$.  In other words, measurements of \slrr\ yield information about the square of the density of quasiparticle states at the Fermi level.  Herein lies the power of NMR to probe the phenomenon of superconductivity: changes to the Fermi surface, such as the development of a gap, will be reflected in \slrr. It is also immediately obvious that  $T_1T K^2$ should be a constant in a Fermi liquid.  This Korringa constant, $\mathcal{K} = \pi^2\hbar \gamma^2/\mu_B^2$, is valid for non-interacting systems, and there are very few experimental systems for which $T_1TK^2/\mathcal{K}$ is unity.  The value of this ratio, in fact, is a measure of the strength of the quasiparticle interactions \cite{PinesCEchi}.

In 1957, Toru Moriya recognized that the expression (\ref{eqn:t1general}) can be rewritten also in terms of the dynamical susceptibility of the quasiparticles.  By employing the fluctuation-dissipation theorem, Eq. (\ref{eqn:t1general}) can be written as:
\begin{equation}
T_1^{-1} =\gamma^2 k_B
T\lim_{\omega\rightarrow 0} \sum_{\mathbf{q}}A^2(\mathbf{q})\frac{\chi''(\mathbf{q},\gamma)}{\hbar\omega},
\label{eqn:t1moriya}
\end{equation}
where the form factor $A^2(\mathbf{q})$ is the square of the Fourier transform of the hyperfine interaction, and the sum over $\mathbf{q}$ is over the first Brillouin zone.  For a contact interaction, $A^2(\mathbf{q})$ is constant, but for more complex situations involving transferred couplings between neighboring sites, $A^2(\mathbf{q})$ can have structure and may vanish at particular wavevectors.  A $\mathbf{q}-$dependent form factor can have profound consequences for the behavior of \slrr\ in materials.  A notorious example is the difference in \slrr\ observed for the planar Cu and planar O in the cuprates \cite{BarrettSlichterYBCO}.  Each nucleus has a different form factor and the dynamical susceptibility of this material is dominated by fluctuations at a particular wavevector, $\mathbf{Q}$.  Since  $A^2(\mathbf{q})$ vanishes for the O site but not for the Cu site, the two \slrrtext s have very different temperature dependences, even though they are coupled to the same degree of freedom, the Cu 3d$^9$ $S=1/2$ spins.

%% file: NMRSChyperfine.tex
\section{Probing the superconducting order via the hyperfine field\label{sec:NMRSChyperfine}}

\subsection{Knight shift measurements in type-I and type-II superconductors}

The hyperfine interaction couples nuclear spins to quasiparticle spins, but in a superconductor at zero temperature, all of the $S=\frac{1}{2}$ electrons are condensed as Cooper pairs in the superconducting condensate. The response of the nuclei depends critically on the spin of the Cooper pairs, which can be either even parity (spin singlet) pairing with net spin $S=0$ or odd parity (spin triplet) with $S=1$. For spin singlet pairing there is no coupling of nuclei to the Cooper pairs, and hence one expects all hyperfine fields to vanish at $T=0$ in the superconducting state.   Therefore, measurements of the Knight shift, or the spin susceptibility, in the superconducting state offer powerful insight into the symmetry of the Cooper pairs.  The spin susceptibility of a superconductor with spin-singlet Cooper pairs is given by the Yosida function:
\begin{equation}
\chi_s(T) = -2\chi_n\frac{T}{T_c}  \int_0^{\infty}\frac{N_s(E)}{N_n(0)}\frac{df(E)}{dE} dE
\label{eqn:yosida}
\end{equation}
where $N_s(E)$ is the density of states in the superconducting state, $f(E)$ is the Fermi-Dirac distribution, and $\chi_n$ is the (Pauli) susceptibility in the normal state. This function clearly vanishes as $T\rightarrow 0$, a result that is independent of the applied field direction.

For spin-triplet pairing, the $S=1$ Cooper pairs do couple to the nuclei, but the hyperfine field depends strongly on the orientation of the applied magnetic field with respect to the spin of the Cooper pairs. In other words, the behavior of the Knight shift below $T_c$ depends on the projection of the $S=1$ Cooper pair spin along $\mathbf{H}$. To illustrate this anisotropy, we note that the superconducting order parameter can be written as: $|\psi(\mathbf{k})\rangle = \Delta_{\downdownarrows}(\mathbf{k})|\downdownarrows\rangle+
\Delta_{0}(\mathbf{k})(|\uparrow\downarrow\rangle+|\uparrow\downarrow\rangle)
+\Delta_{\upuparrows}(\mathbf{k})|\upuparrows\rangle$.  Note here that $|\psi(\mathbf{k})\rangle$ spans the states $\{|S_z=+1\rangle,|S_z=0\rangle,|S_z=-1\rangle\}$ in spin space.  If we transform to a new basis $\{|S_x=0\rangle,|S_y=0\rangle,|S_z=0\rangle\}$, where $\hat{\mathbf{x}}\equiv|S_x=0\rangle = (-|\upuparrows\rangle+|\downdownarrows\rangle)/\sqrt{2}$, $\hat{\mathbf{y}}\equiv|S_y=0\rangle = (|\upuparrows\rangle+|\downdownarrows\rangle)/\sqrt{2}$,
and $\hat{\mathbf{z}}\equiv|S_z=0\rangle = (|\uparrow\downarrow\rangle+|\downarrow\uparrow\rangle)/\sqrt{2}$, then we can write:
$|\psi(\mathbf{k})\rangle = d_x(\mathbf{k})\hat{\mathbf{x}}+d_y(\mathbf{k})\hat{\mathbf{y}}+
d_z(\mathbf{k})\hat{\mathbf{z}}$.  The order parameter is thus defined by a vector, $\mathbf{d}(\mathbf{k})=\frac{(\Delta_{\downdownarrows}-
\Delta_{\upuparrows})}{2}\hat{\mathbf{x}}+
\frac{(\Delta_{\downdownarrows}+
\Delta_{\upuparrows})}{2i}\hat{\mathbf{y}}+\Delta_0\hat{\mathbf{z}}$ in spin space. For the case $\mathbf{d}\times\mathbf{d}^*=\mathbf{0}$ (unitary state), $\mathbf{d}$ defines a direction in spin space given by the normal to a plane in which the spin expectation value remains finite.  Typically, the orientation of the d-vector is pinned to the crystalline axes by the spin-orbit coupling. Depending on the components of $\mathbf{d}$ relative to the applied magnetic field, $\mathbf{H}$, the expectation value of the spin either vanishes or remain finite.   For example, in Sr$_2$RuO$_4$, the order parameter is commonly accepted to be $\mathbf{d}(\mathbf{k}) = \Delta_0\hat{\mathbf{z}}(k_x\pm ik_y)$, which corresponds to $L_z=+1$, with the paired spins lying the the $xy$ plane. In this case, if $H$ lies along the $z$ direction, $K_z$ will vanish since $\langle S_z \rangle=0$, but $K_{xy}$ will remain finite.  Indeed, the Knight shift in the superconducting state of Sr$_2$RuO$_4$ shows exactly such behavior \cite{IshidaSr2RuO4}. For further details on the triplet state in Sr$_2$RuO$_4$ the reader is referred to \cite{mackenziemaenoreview}, and for a more general treatment of the order parameter in a triplet superconductor, see \cite{SigristSCReview}.


Measurement of the Knight shift in a superconductor is a non-trivial enterprise, and may seem contradictory at first glance.  After all, how can one measure a resonance frequency in a material experiencing a Meissner effect?  Furthermore, the magnetic susceptibility of a superconductor is dominated by the diamagnetic response of the orbital current response, and the spin susceptibility is only a small paramagnetic contribution.  In fact, it is precisely because of the hyperfine coupling that one can extract the spin contribution in an NMR experiment, which cannot be done via bulk magnetization, or any other measurements.   However, measurements of the Knight shift in a type-I superconductor are possible only in special situations in which the magnetic field can penetrate the sample, such as in the case of surface superconductivity.  Some of the first NMR experiments on superconductors focused on Knight shift measurements of superconducting Hg colloids, where the surface area is very large \cite{knightshifttype1}.  In a type II superconductor, the situation is much clearer. In this case, the magnetization becomes position dependent in a vortex lattice, where $M(\mathbf{r})= M_{\rm orb}(\mathbf{r}) + \chi_{\rm spin} H$.  Since the nuclei couple to the local internal magnetic induction, the resonance frequency will become position dependent as well: $f(\mathbf{r}) = \gamma (H + 4\pi M(\mathbf{r}) + KH)$.  In a type-I superconductor where $M = -H/4\pi$, the resonance frequency will be essentially zero except in special cases.  However, in a type II superconductor, $|M| < H/4\pi$ and so the resonance frequency is non-zero, although reduced from its value in the normal state.

Experimentally, one often finds the resonance frequency shifts down in the superconducting state, as expected.  The challenge is to extract the Knight shift from the somewhat complex spectra, since the dominant effect may be from the diamagnetic response of the orbital currents, rather than the suppression of spin susceptibility.  The center of gravity of the distribution $\langle M_{\rm orb}\rangle = \int M_{\rm orb}(\mathbf{r}) d\mathbf{r}$ gives rise to a second contribution to the shift of the resonance line: $K_{\rm orb} = \langle M_{\rm orb}\rangle/H$, which is a complex function of field, the penetration depth, and the upper critical field $H_{c2}$.  It is generally impossible to calculate the absolute value of $\langle M_{\rm orb}\rangle$.  Furthermore, there is also a demagnetization factor $N/4\pi$ that shifts the net internal field inside the sample as a result of the diamagnetic magnetization, and suppresses the resonance frequency. This demagnetization field is a strong function of sample geometry, and depends on the direction of the external field with respect to the sample.   The combined effect of $K_{\rm orb}$ and the demagnetization field can cause the resonance frequency to change dramatically in the superconducting state, even in the absence of a Knight shift (spin susceptibility).  In practice, then, it can be particularly challenging to determine what part of the measured shift arises from a suppression of spin susceptibility and what arises because of the diamagnetic response.  However, in superconductors which have more than one NMR-active nucleus, one can sometimes decouple these two effects if one knows the hyperfine coupling to each \cite{CurroPuCoGa5}.   A critical evaluation of any Knight shift measurement in the superconducting state needs to consider all of these effects in detail.
A further challenge to measurements of the Knight shift at $T=0$ is the spin-orbit coupling. Spin-orbit coupling mixes spin-up and spin-down states such that the Cooper pairs are composed of a mixture of the two, and the ground-state spin susceptibility will approach a finite value at $T=0$ rather than vanish. The review by MacLaughlin discusses these effects in detail, as well as several experimental results \cite{maclaughlinSCreviewbook}.

Once the zero-temperature Knight shift is determined, and the orbital, demagnetization field, and spin-orbit effects have been evaluated, the remaining temperature dependent shift arises from the suppression of the spin susceptibility in the superconducting state.  In principle, the detailed temperature dependence can be calculated with Eq. (\ref{eqn:yosida}), and compared with the density of states to extract quantities such as the magnitude of the superconducting gap, $|\Delta|$, and the symmetry of the gap in $\mathbf{k}$-space.  In practice, however, the differences between the calculated shifts for different gap functions easily can fall within the noise.

It should be noted that the SNR in the superconducting state is usually significantly reduced as a result of the enhanced screening of the rf excitation pulses.  Whereas in the normal state, the rf skin depth is on the order of a few microns ($B(r) \sim e^{-(1+i)r/\delta}$), in the superconducting state, the rf penetration depth is often only an order of magnitude less  ($B(r) \sim e^{-r/\lambda_C}$) \cite{CoffeyClem}.  For NMR Knight shift measurements, one requires single crystals rather than powders, and therefore this change of NMR-active volume can significantly reduce the signal, particularly just below $T_c$.  As the temperature is lowered below $T_c$, the Boltzmann factor will tend to enhance the signal.  In a type-II superconductor, the rf pulses penetrate to within the Campbell depth, $\lambda_C > \lambda$. In effect, the vortex lattice itself responds to the external time-dependent field, resulting in a greater rf penetration.

\subsection{Spin-lattice relaxation in superconductors}

 In a fully gapped spin-singlet superconductor at $T=0$ there are no fluctuations of $H_{\rm hyp}(t)$ to couple to the nuclei.  At finite temperature, quasiparticle excitations couple to the nuclear spins, but since the excitations of a BCS state are actually superpositions of spin-up and spin-down quasiparticles, the hyperfine field experienced by the nuclei is different than in the normal state.  This fact has profound consequences for the \slrrtext\ in superconductors, and gives rise to the Hebel-Slichter coherence peak.  To see this, we can write Eq. (\ref{eqn:t1general}) as:
 \begin{equation}
 T_{1}^{-1} \propto A^2 \int
c^{\dagger}_{\mathbf{k}'\downarrow}c_{\mathbf{k}\uparrow}d\mathbf{k}d\mathbf{k}',
\label{eqn:T1normalstate}
 \end{equation}
where $c^{\dagger}_{\mathbf{k}s}$ is the creation operator for the state $|\mathbf{k}s\rangle$. This expression clearly describes a spin-flip scattering event weighted by the square of the hyperfine coupling. The  Buglioubov excitations in a superconductor are coherent superpositions of the states $|\mathbf{k}\uparrow\rangle$ and $|-\mathbf{k}\downarrow\rangle$, such that:
\begin{equation}
\alpha^{\dagger}_{\mathbf{k}1,2}=u_{\mathbf{k}}c^{\dagger}_{\mathbf{k}\uparrow}\pm v_{-\mathbf{k}}c_{-\mathbf{k}\downarrow},
\label{eqn:bugoliubov}
\end{equation}
 where $\alpha^{\dagger}_{\mathbf{k}1,2}$ is the creation operator of a Bugoliubov excitation above the superconducting condensate, and $u_{\mathbf{k}}$ and $v_{\mathbf{k}}$ are their respective amplitudes.  Therefore, \slrr\ can be rewritten as:
 \begin{equation}
 T_{1}^{-1} \propto A^2 \int(u_{\mathbf{k}'}u_{\mathbf{k}} +
v_{\mathbf{k}'}v_{\mathbf{k}})\alpha^{\dagger}_{\mathbf{k}'1}\alpha_{\mathbf{k}2}
d\mathbf{k}d\mathbf{k}'.
\label{eqn:T1SCstate}
 \end{equation}
 This expression shows that the spin-lattice relaxation in the superconducting state is driven by spin-flip scattering of Bugoliubov excitations, but weighted by a coherence factor $C_+(\mathbf{k},\mathbf{k'}) = u_{\mathbf{k}'}u_{\mathbf{k}} +
v_{\mathbf{k}'}v_{\mathbf{k}}$, a quantity that is determined by the superconducting condensate parameters.  If the superconducting state were actually a simple two-fluid system rather than composed of coherent Cooper pairs, then the relative phase of $v_{\mathbf{k}}$ with respect to $u_{\mathbf{k}}$ would vanish, $C_+(\mathbf{k},\mathbf{k'})$ would average to unity, and Eq. (\ref{eqn:T1SCstate}) would simply describe relaxation by a new fluid of $\alpha^{\dagger}_{\mathbf{k}n}$ quasiparticles.  However, since the  Bugoliubov excitations are indeed \textit{coherent} superpositions of the simultaneous spin-up and spin-down electrons, the coherence factor is greater than unity, which has profound consequences for the relaxation rate, as shown below.

\begin{figure}
\centering
\includegraphics[width=\textwidth]{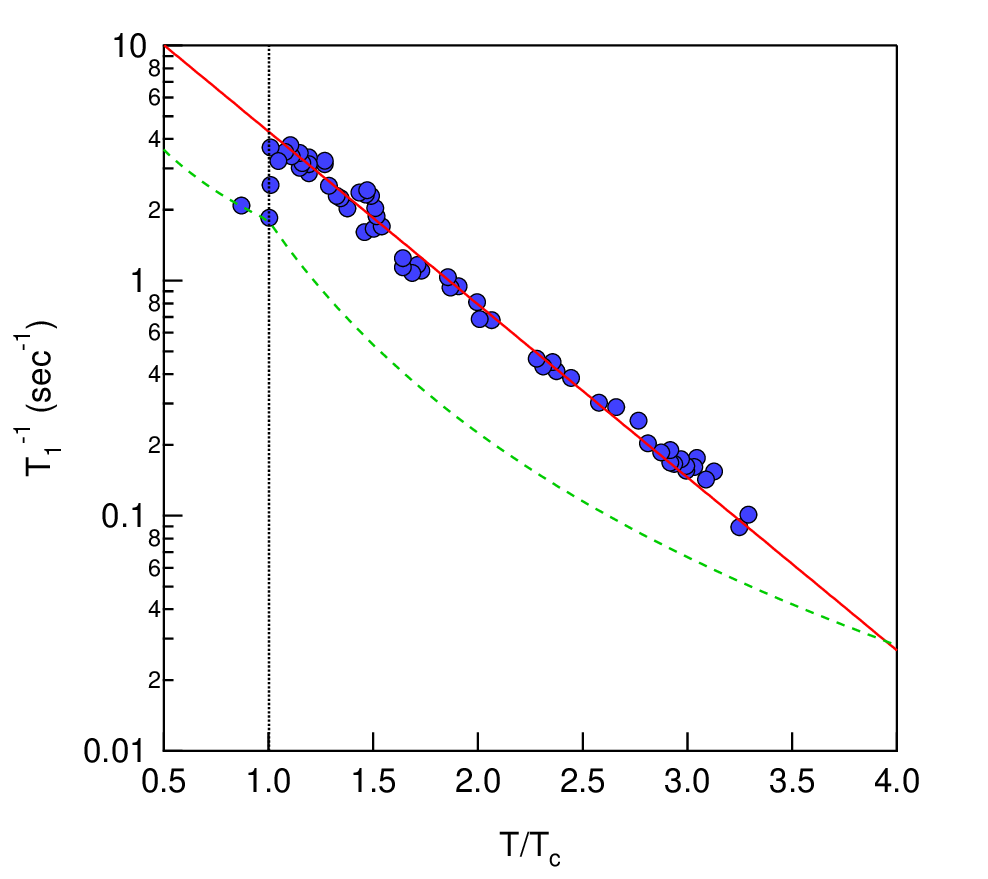}
\caption{The relaxation rate of the $^{27}$Al in the normal and superconducting states, reproduced from
\cite{masudaredfield}.  Note the dramatic coherence peak in \slrr\ just below $T_c$.  The solid red line is a fit to $T_1^{-1} \sim \exp(-\Delta/k_BT)$ as described in the text, with $2\Delta = 3.4 k_BT_c$, and the dashed green line is a plot of $T_1^{-1}  \sim T^3$, as expected for a nodal superconductor.  \label{fig:redfieldT1}}
\end{figure}

\begin{figure}
\centering
\includegraphics[width=\textwidth]{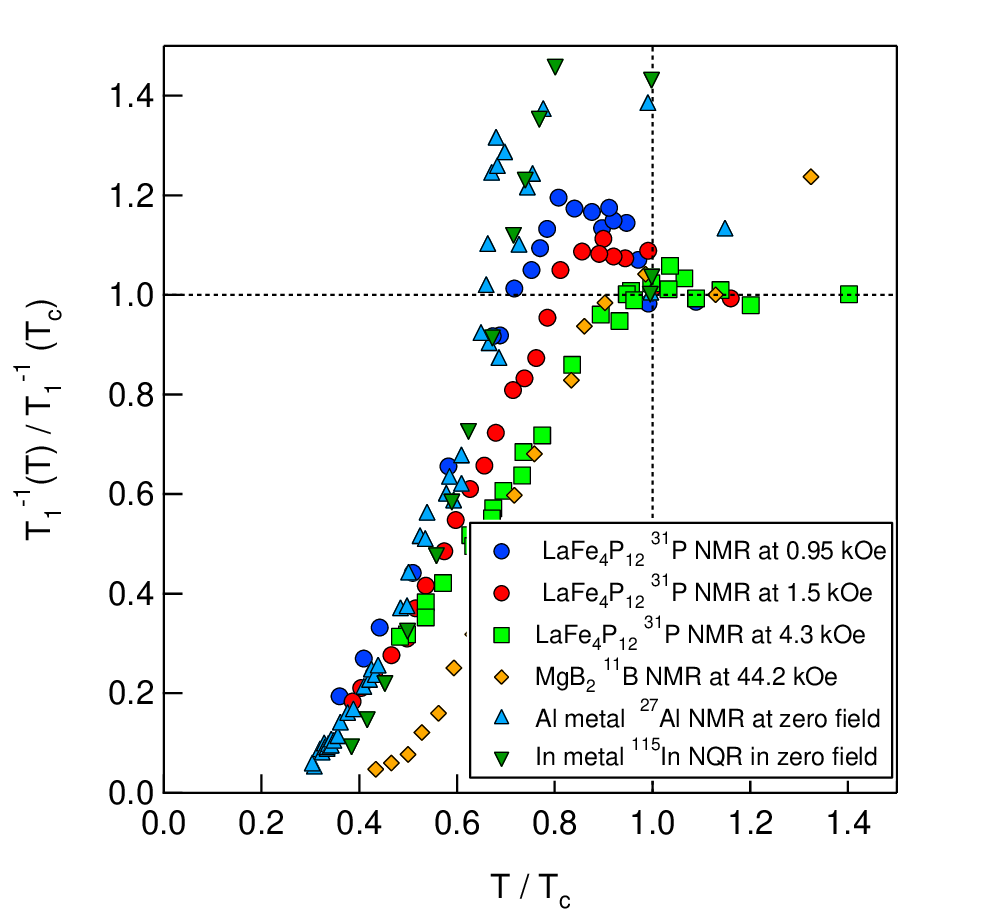}
\caption{The relaxation rates of the skutterudite compound LaFe$_4$P$_{12}$ \cite{IshidaLaFe4P12NMR}, for MgB$_2$ \cite{MgB2NMR}, elemental Al  \cite{masudaredfield}, and elemental In \cite{superconductingInT1}. These  materials are all s-wave superconductors, yet the visibility of the coherence peak varies dramatically. \label{fig:coherencecomparison}}
\end{figure}

Eq. (\ref{eqn:T1SCstate}) can be simplified by noting that the number of Bugoliubov quasiparticles is given by the Fermi-Dirac distribution, and employing the superconducting state density of states:
\begin{equation}
T_1^{-1} \propto A^2 \int C_+(E,E')N_{\rm BCS}(E)f(E)N_{\rm BCS}(E')(1-f(E')) dE,
\label{eqn:t1SCgeneral}
\end{equation}
 where the density of states of an s-wave ($L=0$ symmetry of the Cooper pairs) is given by:
\begin{equation}
N_{\rm BCS}(E) =\frac{E}{\sqrt{E^2-\Delta^2}}N_{\rm N}(E)
\end{equation}
and the coherence factor is:
\begin{equation}
C_+(E,E') =\frac{1}{2}+\frac{E E'}{2\Delta^2}.
\end{equation}
Since $N_{\rm BCS}(E)$ has a singularity at $E=\Delta$ and $C_+(E,E')$ is non-zero, expression (\ref{eqn:t1SCgeneral}) leads to a singularity of \slrr at $T=T_c$.  This is the Hebel-Slichter coherence peak, first observed experimentally in superconducting Al as an unexpected peak in \slrr\ just below \tc\ \cite{hebelslichter2}.  The original data measured by Masuda and Redfield is shown in Fig. \ref{fig:redfieldT1}, which clearly shows the dramatic increase in \slrr\ just below $T_c$. The intensity of the peak is limited in practice by both a smearing of the density of states as well as spin-orbit scattering effects.  The spin-orbit interaction mixes spin-up and spin-down states with different $|\mathbf{k}\rangle$ states, so the Bugoliubov states in Eq. \ref{eqn:bugoliubov} are modified to include a sum over $\mathbf{k}$ states, as well as  the coherence factor $C_+(\mathbf{k},\mathbf{k'})$.  In practice, the coherence factor can be rather difficult to observe experimentally, particularly in materials with heavier elements where the spin-orbit interaction is larger. Fig. (\ref{fig:coherencecomparison}) shows data for a number of s-wave superconductors, and clearly shows how the intensity of the coherence peak can be strongly suppressed, particularly in larger fields. This suppression may be due in part to localized excitations within the vortex cores \cite{ichioka}. The reader is referred to \cite{maclaughlinSCreviewbook} for further details. It is important to note that this peak is relatively hard to see in most materials and that its absence is not
necessarily an indication of unconventional superconductivity.

At this point, it is worth noting that the acquisition of these data during the time period 1957 - 1961 represents an experimental \textit{tour de force}, since Al is a type-I superconductor with a transition temperature of 1.17 K.  Although other nuclei were available with higher $T_c$'s, or even $^{115}$In with an NQR frequency of 7.54 MHz and $T_c = 3.41$ K, Al was chosen because the low Z reduced potential effects of spin-orbit coupling (In NQR was performed ten years later by Butterworth and MacLaughlin \cite{superconductingInT1}). Hebel and Slichter  at the University of Illinois and Masuda and Redfield at IBM both tackled the problem independently by constructing an ingenious field cycling technique to let the nuclei relax in zero external field (in the superconducting state), while measuring the nuclear magnetization in an applied field in the normal state \cite{masudaredfield,hebelslichter2}.  Hebel and Slichter were able to reach temperatures $\sim 0.94$ K by pumping on liquid He-4.  Masuda and Redfield reached 0.35 K by pumping on He-3 using a special He-3 cryostat that had just been developed \cite{He3cryostat}. Aside from the rigorous and technically challenging field and temperature requirements, even simple NMR measurements were challenging by today's standards, where standard NMR spectrometers consist of computer controlled digital radio transmitters and receivers. Indeed, many frequency measurements were made by using a Lecher-wire assembly, in which radiofrequencies were measured by setting up standing waves on parallel wires, and the nodes of the standing waves were measured by connecting a lightbulb across the pair at various distances.

For $T\lesssim T_c/3$, Eq. (\ref{eqn:t1SCgeneral}) reduces to $T_1^{-1} \sim \exp(-\Delta/k_BT)$.  In other words, the \slrrtext\ exhibits an activated temperature dependence, as expected in the presence of a uniform gap function.  This behavior, however, is only present if the gap is isotropic in $\mathbf{k}$-space, i.e., for s-wave superconductivity.  If $L>0$ and $\Delta_{\mathbf{k}}$ develops nodes, then the density of states $N_{\rm BCS}(E)$ is modified.  For line nodes in an $L=2$ (d-wave) superconductor, $N_{\rm BCS}(E)\sim E/\Delta_0$ for $E\ll\Delta_0$, where $\Delta_0$ is the maximum gap value.  In this case, Eq.(\ref{eqn:t1SCgeneral}) leads to $T_1^{-1} \sim T^3$ for $T \lesssim T_c/3$. For point nodes, as may be the case for $L=1$ (triplet) pairing, then $T_1^{-1} \sim T^5$ \cite{SigristSCReview}. These dramatic differences in temperature dependence arises because there are always quasiparticle excitations available at the gap nodes.  Measurements of the temperature dependence of \slrr\ can help to distinguish between these different types of pairing.  It should be noted, however, that \slrr\ is only sensitive to the magnitude of of the gap, not the sign.  In principle, if the gap is anisotropic s-wave, then the density of states can mimic that of a true superconductor with nodes in the gap.  In the case of the cuprates, the quantity $T_{2G}$, the Gaussian component of the spin-spin relaxation rate, is sensitive to the change of sign of the gap function in $\mathbf{k}$-space.  $T_{2G}$ is a measure of the indirect coupling between nuclei as mediated by the real part of the electronic spin susceptibility, $\chi'(\mathbf{q},\omega=0)$.  This quantity is modified below \tc\ and is a function of $\Delta_{\mathbf{k}\pm\mathbf{q}}$.  Since $T_{2G}$ involves the coupling of two different nuclei to $\chi'(\mathbf{q},\omega=0)$, it can distinguish between an anisotropic s-wave and a true d-wave superconductor \cite{wrobel}.  Indeed, measurements in YBa$_2$Cu$_4$O$_8$ are consistent with calculations based on d-wave, rather than anisotropic s-wave, superconductivity \cite{Corey1248}.  This connection between $T_{2G}$ and $\chi'(\mathbf{q},\omega=0)$ is a special case that happens to apply to the Cu nuclei in the cuprates. The hyperfine couplings of the Cu are such that the \textit{indirect} nuclear-nuclear coupling is Ising-like with a particular direction (the $c$ direction for the Cu).  In this case, the form of the echo decay is Gaussian, and is characterized by the quantity $T_{2G}$, which can be directly related to $\chi'(\mathbf{q},\omega=0)$ \cite{CurroSlichterEchoDecay}. In general, though, the indirect couplings are weaker and more isotropic than in the cuprates, and consequently there is little information to be extracted from the form of the echo decay.

\section{Probing the vortex lattice\label{sec:NMRSCorbital}}

\subsection{Field distributions}
\begin{figure}
\centering
\includegraphics[width=\textwidth]{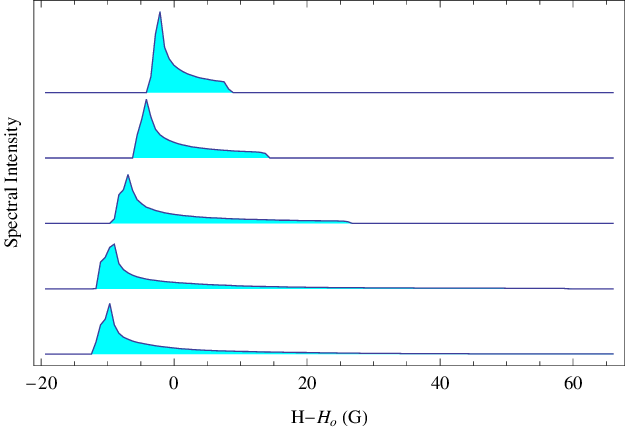}
\caption{The field distribution $\mathcal{P}(h)$ versus local field $h=H-H_0$ in a series of external fields $H_0=$ 0.5, 1.0, 5.0, 10.0 and 15.0 T  for a superconductor with $H_{c2}=36.0$ T and penetration depth $\lambda = 2700$ {\AA}. The different fields have been offset vertically (with increasing field) for clarity. \label{fig:redfield}}
\end{figure}

In a type II superconductor, the inhomogeneous distribution of supercurrent in a vortex lattice gives rise to a periodic array of local \textit{orbital} magnetization, $M_{\rm orb}(\mathbf{r})$.  The orbital magnetization in a vortex lattice couples to the nuclear spins via a chemical shift: $\mathcal{H}_{\rm orb} = 4\pi\mathbf{\hat{I}}\cdot\mathbf{M}_{\rm orb}(\mathbf{r})$.  Typically a chemical shift is given by an on-site diamagnetic orbital susceptibility in a solid or molecule.  In a superconductor, the diamagnetic response of the supercurrents is a macroscopic phenomenon, with an inhomogeneity set by the scale of the magnetic field.  The unit cell of a triangular vortex lattice is given by: $r_{VL} \sim \sqrt{\Phi_0/H_0} \sim \sqrt{H_{\rm c2}/H}\xi$, where $\xi$ is the coherence length.  Since $r_{VL} \gg a$, the unit cell length, the local orbital shift (and hence the NMR frequency) varies on a length scale much larger than the vortex lattice.  Each nucleus resonates at the local magnetic field $f(\mathbf{r}) = \gamma|H_0 + 4\pi \mathbf{M}_{\rm orb}(\mathbf{r})|$, so the NMR spectrum consists of a histogram of local fields in the vortex lattice, $\mathcal{P}(f) \sim 1/|\nabla f(\mathbf{r})|$ (see Fig. \ref{fig:redfield}). This characteristic spectrum, the "Redfield pattern", has been used extensively to determine properties of the vortex lattice in a number of superconductors \cite{brandt,SonierReview,maclaughlinSCreviewbook}. Ideally, one can measure the spectrum $\mathcal{P}(f)$ and fit the data to an analytic expression, but this is not realistic in practice.  Typically other inhomogeneities limit the spectral resolution such as the magnet inhomogeneity. Fig. \ref{fig:ishidaLaRu4P12} shows the $^{31}$P spectrum in the mixed state of LaRu$_4$P$_{12}$ \cite{ishidaLaRu4P12}.  Although the spectrum is broadened, a clear tail is evident at the upper frequency, corresponding the the classic Redfield pattern.

As discussed above, there can be errors in determining the absolute value of the shift, given by the first moment of the spectrum, because there are several quantities that affect the resonance frequency in the superconducting state. The linewidth, or the rms second moment of the spectrum, is often more useful and less prone to extrinsic effects. The second moment is given by
\begin{equation}
\langle (f/\gamma -H_0)^2 \rangle \approx 0.00371 \frac{\Phi_0^2}{\lambda^4}
\label{eqn:VLmomentlowfield}
\end{equation}
for $H\ll H_{c2}$ and
\begin{equation}
\langle (f/\gamma- H_0)^2 \rangle \approx (7.52\times 10^{-4}) \frac{\Phi^2}{\lambda^4}\frac{\kappa^4(1-H/H_{c2})^2}{(\kappa^2-0.069)^2}
\label{eqn:VLmomenthighfield}
\end{equation}
 for $H$ close to $H_{c2}$, where $\lambda$ is the penetration depth, and $\kappa = \lambda/\xi$ is the Ginsburg-Landau ratio. Therefore, measurements of the lineshape can provide direct information about the superconducting penetration depth (and hence the superconducting condensate density $n_s \sim 1/\lambda^2$) as a function of temperature \cite{CurroPuCoGa5,SonierReview}.

Vortex matter can undergo phase transitions, and NMR can probe the physics of the vortex lattice phase diagram.  In two-dimensional layered systems such as the high temperature superconductors, the pancake vortices can be either coupled or decoupled between layers, and undergo a vortex liquid to solid transition \cite{blatter}. Since the vortices can move in the liquid state, they create a dynamic field $\delta\mathbf{M}(\mathbf{r},t)$ at the nuclear site that can lead to relaxation.  If the autocorrelation time of the longitudinal fluctuations, $\tau$, defined as $\langle \delta M_z(\mathbf{r},t)\delta M_z(\mathbf{r},0)\rangle\sim e^{-t/\tau}$, is short compared to $1/\gamma\delta M_z$, then the spectrum will be motionally narrowed, and will not show the characteristic Redfield pattern. When the vortex lattice cools and undergos a liquid to solid transition, the correlation time $\tau$ will become much longer, in which case the condition $1\gamma\delta M_z\tau \gg 1$ will be satisfied. In this case, the spectrum will no longer be motionally narrowed, and the Redfield pattern will emerge.  This technique was applied to map the vortex lattice phase diagram in YBa$_2$Cu$_3$O$_7$ \cite{ReyesYBCOvortexmelting}.  Also, in many instances anomalies have been found in \slrr\ and $T_2^{-1}$ at characteristic temperatures inside the superconducting state, which are probably related to motions of the vortex lattice \cite{penningtonT2YBCO,Corey1248}.  A solid vortex lattice can also experience overdamped vibrational modes as a results of the finite moduli in the lattice \cite{brandtmoduli,blatter}.  These modes can give rise to fluctuating fields at the nuclear sites close to the vortex cores, and may contribute a significant portion to the local \slrrtext.  For this reason, it is advantageous to measure \slrr\ at zero field, in the absence of vortices.  Typically this is done in an NQR experiment, if a quadrupolar nucleus is available and has a reasonably large quadrupolar frequency ($> 1$ MHz).

\subsection{Interaction with radiofrequency irradiation}
The radiofrequency NMR pulses used for exciting the resonant nuclei can also drive motion of the vortex lattice.  In order to investigate the effects of the rf pulses on the vortex lattice itself, researchers at UCLA have made systematic studies of how the rf pulses can anneal a distorted vortex lattice \cite{clarkRIFLA}. In these studies, a type-II superconducting material is field-cooled, then rotated by an angle on the order of a few degrees with respect to the static field.  This rotation strains the vortex lattice, giving rise to a large inhomogeneous field distribution.  In a standard spin-echo experiment, the field inhomogeneity is refocused, and the original signal is recovered at a time $2\tau$ from the first pulse.  In the case of the strained vortex lattice, the magnitude of the echo is reduced from that of the unstrained lattice.  The reason is the refocusing $\pi$ pulse partially anneals the vortex lattice, and changes the local field at points in the vortex lattice where the positions of the vortex lines have changed.  In these regions, the signal is not refocused.  However, after a series of pulses, the full signal intensity can be recovered.  This experiment suggests that the rf pulses used in NMR of type II superconductors \textit{do not significantly affect} the equilibrium field distribution created by field-cooled conditions.

The rf pulse annealing experiments also suggest that significant field distributions can exist when the vortex lattice is strained. Therefore, NMR experiments that are designed to probe a field distribution (such as measurements of the penetration depth described above) should \textit{always} be conducted under field-cooled conditions.  However, in order to extract an NMR spectrum, it is often easier and more straightforward to take data by sweeping frequency at constant field rather than take data by sweeping frequency at constant field.  Obviously the latter is going to be more accurate.  Indeed, the former technique is likely to reveal extrinsic details of the field inhomogeneity rather than intrinsic details of the physics of the vortex lattice.  If the field is swept inside the Abrikosov vortex lattice state, then the vortices will either enter or depart the sample from the ends, and the magnetization will be inhomogeneous on a macroscopic scale with a magnitude determined by the critical current density, $J_c$, as described in the Bean model \cite{tinkham}.  In this case, it is not straightforward to determine if quantities such as the second moment of the NMR line are probing intrinsic properties such as the penetration depth, or extrinsic properties that depend on the sample geometry.  Unfortunately this phenomenon is not widely recognized or discussed in the NMR community, and non-specialists are cautioned that reports of these types of experiments should always include a careful discussion of the experimental conditions.

\subsection{Imaging experiments in vortex lattice field gradients}

\begin{figure}
\centering
\includegraphics[width=\textwidth]{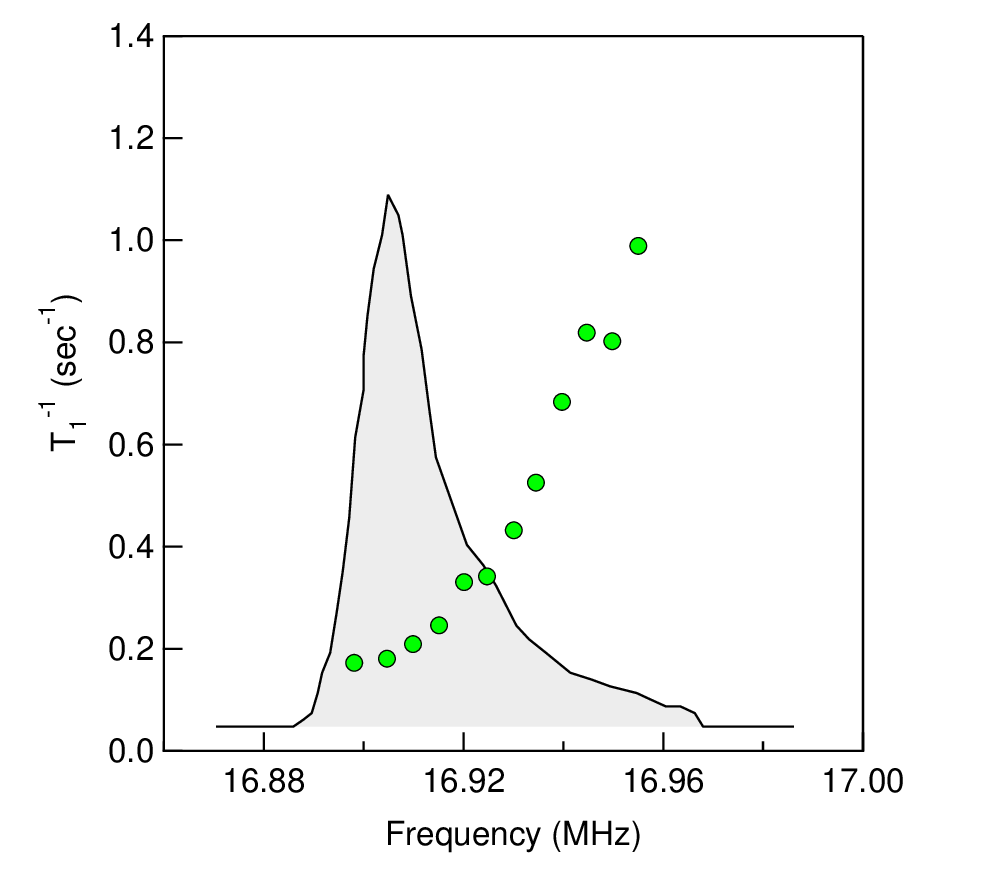}
\caption{The $^{31}$P spectrum (solid line) and the \slrrtext\ ($\bullet$) in the mixed state of superconducting LaRu$_4$P$_{12}$ ($T_c=7.2$ K) at 2 K and 5.2 T. Reproduced from \cite{ishidaLaRu4P12}. \label{fig:ishidaLaRu4P12}}
\end{figure}

A significant advantage of NMR experiments in a vortex lattice is the possibility of real-space imaging because of the large field gradients intrinsically present in a vortex lattice.  Since the resonance frequency in the Redfield pattern corresponds to a particular location in real space relative to vortex cores, one can selectively measure the relaxation rates at different parts of the vortex lattice \cite{curroslichter,mitrovicYBCO,kumagaiAFVortex}.  This technique is particularly enlightening in the case of d-wave superconductors, where the local density of quasiparticle states varies spatially because of the so-called "Doppler" shift.  The energy of a quasiparticle excitation which is moving as part of a supercurrent will have an energy:
\begin{equation}
E_{\mathbf k} = \sqrt{\xi_{\mathbf k}^2 + \Delta_{\mathbf k}^2} + \hbar \mathbf{v}_s\cdot\mathbf{k},
\label{eqn:doppler}
\end{equation}
where $\xi_{\mathbf{k}}$ is the normal state dispersion, and $\mathbf{v}_s$ is the local supercurrent velocity. The second "Doppler shift" term was first recognized by Volovik \cite{volovikDOS}, and is important because it gives rise to an enhancement of the local density of states.  At $T=0$ the point nodes of a nodal superconductor become finite arcs in $\mathbf{k}-$space because of this Doppler shift.  Since the supercurrent velocity varies spatially in a vortex lattice with a periodicity commensurate with the local field, there is a correspondence between the local resonance frequency and the local density of quasiparticle states, and hence the \slrrtext.  Several groups have taken advantage of this result to measure the properties of the vortex lattice in the cuprates \cite{curroslichter,mitrovicYBCO,kumagaiAFVortex,wortis,morr,ichioka}.  These experiments generally find a strong spatial dependence of \slrr\ outside of the vortex cores, as predicted.  At short distances, essentially within the vortex core, \slrr\ exhibits very different behavior in the case of YBa$_2$Cu$_3$O$_7$ in high fields, and in Tl$_2$Ba$_2$CuO$_{6+\delta}$ compounds.  This behavior inside the vortex cores (within a radius $\xi$ of the center of the vortices, where $\xi$ is the coherence length) is suggestive of antiferromagnetic fluctuations, and has been taken as evidence for antiferromagnetic vortex cores in the cuprates.  Subsequent inelastic neutron scattering data seem to support this scenario \cite{mitrovic,lake}.  The coexistence of two order parameters is possible since the superconducting order is locally suppressed within the vortex cores.  A similar phenomenon is probably at play in the heavy fermion superconductor \cecoin\ \cite{CurroCeCoIn5FFLO}.

Measurements of the local density of states in an s-wave superconductor can also reveal strong local spatial variations.  Recently, Ishida et al. measured the Redfield pattern in the skutterudite superconductor LaRu$_4$P$_{12}$ ($T_c=$7.2 K), which is an s-wave superconductor.  Fig. \ref{fig:ishidaLaRu4P12} shows a strong frequency dependence of \slrr\ similar to that observed in d-wave superconductors \cite{curroslichter}.  Clearly, the strong spatial dependence in this case cannot be attributed to the Doppler shift (\ref{eqn:doppler}), but must arise either from localized bound states in the vortex cores, or from vortex motion.

%% file: NMRSCheavyelectron.tex
\section{NMR in the Heavy Electron Materials\label{sec:NMRheavy}}

\begin{figure}[!ht]
\begin{center}
\includegraphics[width=\textwidth]{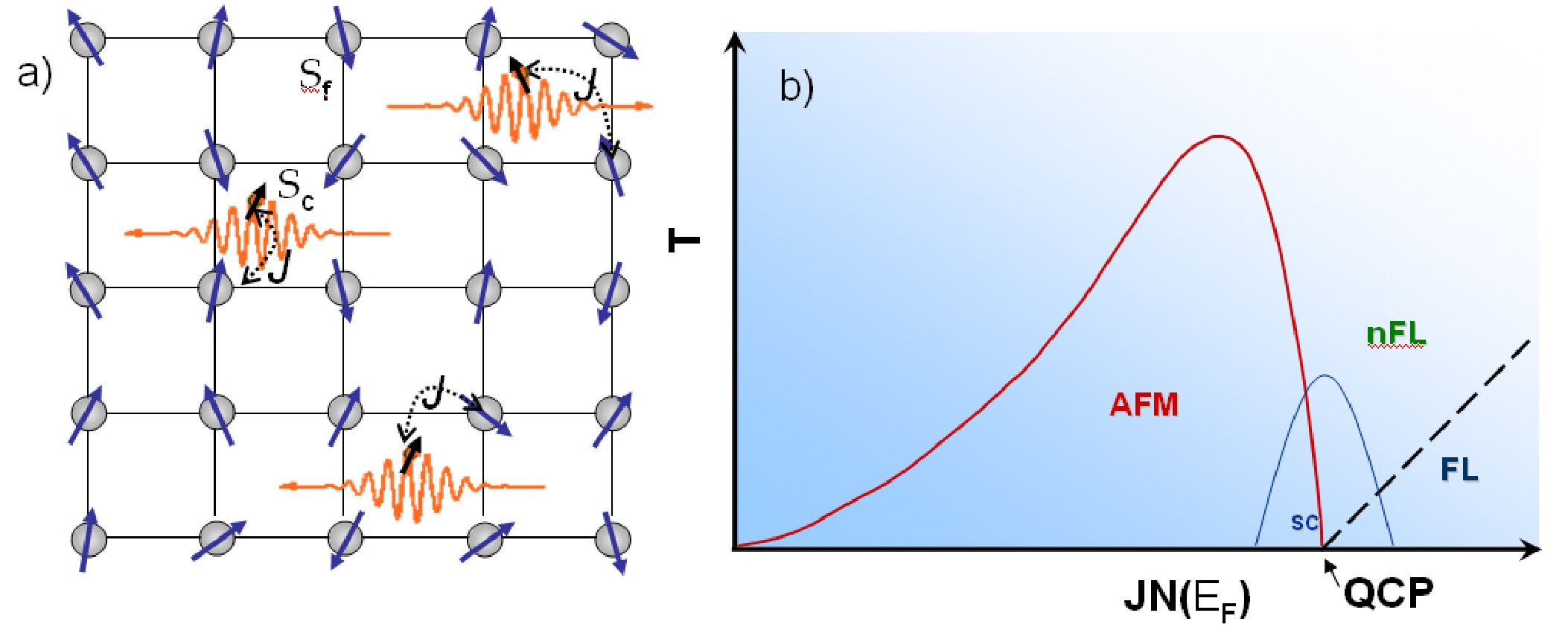}
\caption{ (a) In the Kondo lattice model  a lattice of nearly
localized spins ($S_f$) interact with a sea of conduction electrons
($S_c$), via a Kondo coupling ($J$). (b) The phase diagram of the
Kondo lattice, $T$ versus $JN(0)$, where $N(0)$ is the density of
conduction electron states.} \label{fig:kondolattice}
\end{center}
\end{figure}

Superconductivity in the heavy fermion materials tends to emerge only under certain structures, and in particular regions of the phase diagram.
In the case of the heavy fermions, much of the relevant physics can be captured by
the Kondo lattice model in which a lattice of f-electron moments is
embedded into a background of conduction electrons (Fig.
\ref{fig:kondolattice}a) \cite{doniachbook}. The nature of the
ground state is highly sensitive to the magnitude of the Kondo
exchange interaction, $J$, and the conduction electron density of
states, $N(0)$.  When $J$ and $N(0)$ are small, the f-moments can polarize the
conduction electron medium, leading to an effective RKKY interaction
between the moments and hence a magnetically ordered ground state.
On the other hand, when $J$ and $N(0)$ are large, the conduction electrons form
singlets with the f-moments, reducing the effective moment and
leading to a spin-liquid ground state with no long range order.
Between these two extremes there is a quantum phase transition between an ordered and a disordered state (Fig. \ref{fig:kondolattice}b). New broken symmetries often emerge in the vicinity of this quantum critical point (QCP) where the RKKY and
Kondo interactions compete.

Quantum phase transitions have been focus of much theoretical and
experimental interest in the past decade
\cite{SiLocalQCP,ColemanHFdeath,ColemanQCreview,lonzarichCeIn3,YRSnature}.
The critical fluctuations of a quantum phase transition are driven
by quantum rather than thermal effects, and the observed behavior of
many bulk observables near a QCP deviates strongly from conventional
theories of phase transitions \cite{goldenfeldbook,sachdevbook}. The
Kondo lattice QCP described above is responsible for much of the
new physics and emergent behavior exhibited by heavy fermion
compounds. In these systems, the Kondo interaction, $J \propto V^2$,
is a strong function of the hybridization, $V$, between the
conduction electron and $f$-electron wavefunctions. This
hybridization can be modified by hydrostatic pressure or chemical
doping and hence the ground state can be tuned over a broad
range of phase space (Fig. \ref{fig:kondolattice}b). Consequently,
 heavy fermion materials offer an unprecedented testing ground to investigate the emergence of non-Fermi liquid behavior and the
properties of novel ground states in the vicinity of a QCP.
Although these systems are metallic, observed bulk properties such
as the resistivity and specific heat disagree with the predictions
of Fermi liquid theory, which is normally quite robust. This
breakdown of Fermi liquid theory represents a significant challenge
to theory and has been the subject of extensive experimental work in
recent years \cite{lonzarichCeIn3,YRSnature}.

NMR measurements can play a central role in probing the low energy degrees of freedom that emerge in the vicinity of a QCP. As the ground state evolves from an ordered magnetic state to a disordered state, the excitations above this ground state should change dramatically. These changes should be reflected in the low energy spin fluctuations, and hence in quantities such as \slrr.  Indeed, different theoretical models for quantum phase transitions can make specific predictions for the behavior of \slrr\ \cite{uedareview}. The theory of local quantum criticality developed by Si and coworkers explains the unusual dynamical susceptibility, $\chi(\mathbf{q},\omega,T)$ measured in some compounds, and predicts \slrr$\sim$ constant for sufficiently low temperatures in two dimensions \cite{SiLocalQCP}. The subsequent observation of exactly this behavior by Si NMR in YbRh$_2$Si$_2$ has helped identify this compound as a prime candidate for this theory \cite{IshidaYRS}.

\begin{figure}
\centering
\includegraphics[width=4.0in]{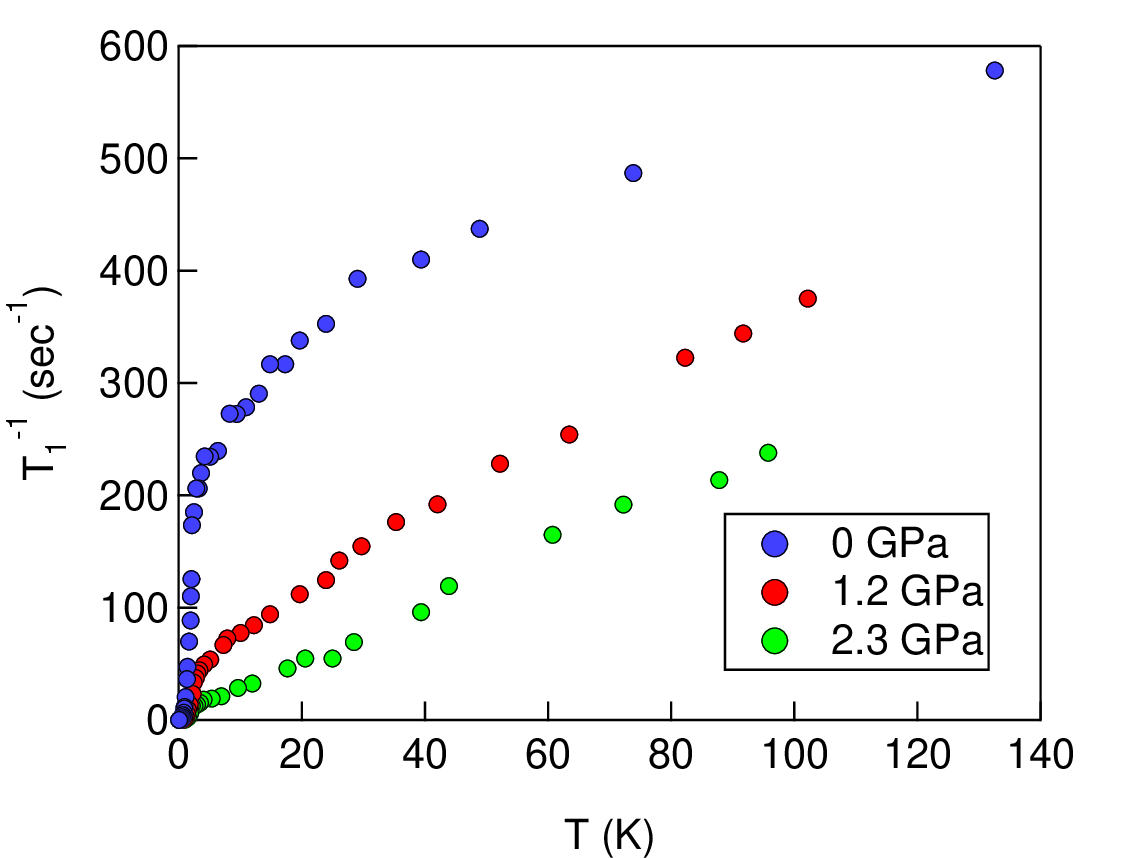}
\caption{\slrr\ versus temperature at various pressures in \cecoin, reproduced from \cite{kohoriCeCoIn5pressure}.  The strong pressure dependence may reflect the evolution of the low energy spin degrees of freedom as the system is tuned away from a QCP near P = 0 GPa.  \label{fig:kohoriT1pressureCeCoIn5} }
\end{figure}

On the other hand, the situation is not so clear in \cecoin, which is a particularly important material because it is a superconductor, and lies close to a QCP at ambient pressure \cite{sidorov}.  Indeed, NMR measurements of the \slrrtext\ as a function of pressure clearly show dramatic changes in the low energy degrees of freedom that are expected in the vicinity of a QCP (see Fig. \ref{fig:kohoriT1pressureCeCoIn5}) \cite{kohoriCeCoIn5pressure}.  The unusual $T^{1/4}$ behavior observed at ambient pressure has been seen as evidence of quantum critical fluctuations \cite{uedareview,kohoriCeCoIn5pressure}.  Although the strong pressure dependence must reflect the evolution away from a QCP, it is not clear whether the changes could be related to changes in critical fluctuations that persist up to 100 K, or perhaps changes in the hyperfine coupling itself.  This material is just one of many examples in which the behavior in the normal state presents several challenges.

\subsection{The Knight shift anomaly}

\begin{figure}
\centering
\includegraphics[width=\textwidth]{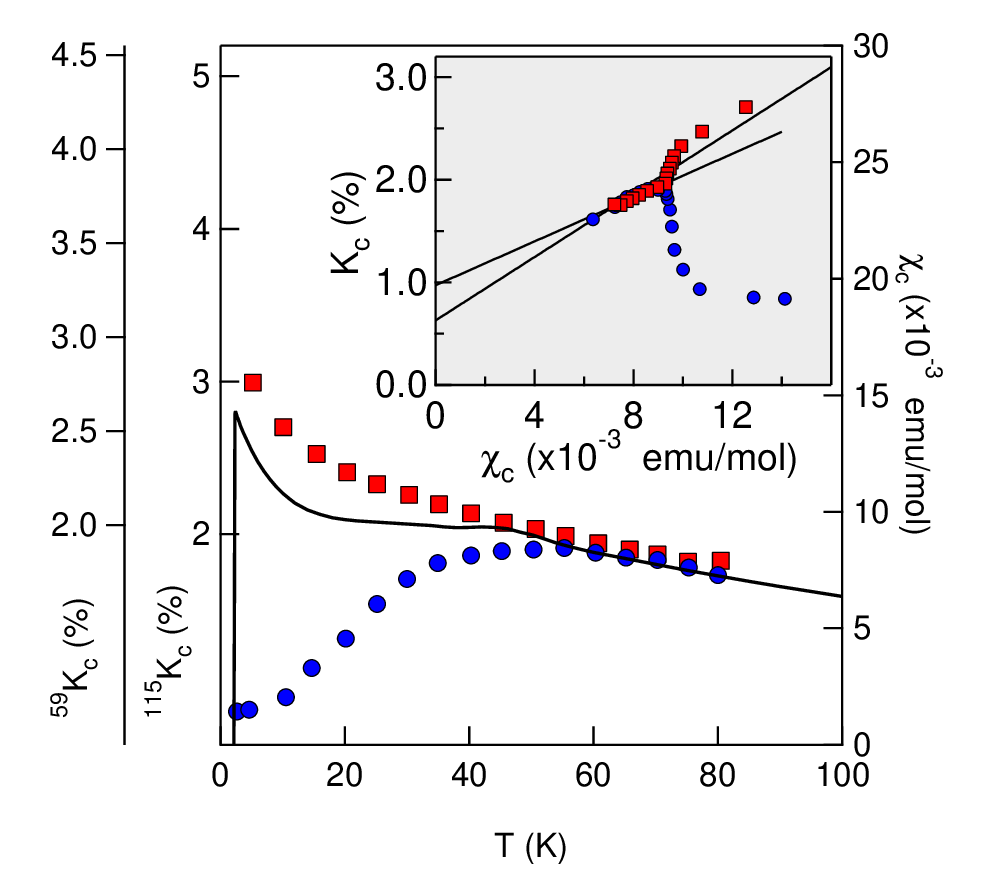}
\caption{The Knight shift of the In(1) (red squares) and the Co (blue circles) in \cecoin, compared with the bulk susceptibility, $\chi(T)$ (solid line), in the $c-$direction.  INSET: $K_c$(In(1)) and $K_c$(Co) versus $\chi_c$ with temperature as an implicit parameter.  The solid lines are linear fits to the data for $T>T^*$.  $T^*$, the temperature where $K$ and $\chi$ diverge, is approximately 50 K in this case. \label{fig:cecoin5shift} }
\end{figure}

In the normal state of heavy fermion materials, the magnetic susceptibility $\chi$ usually increases strongly with decreasing temperature. In many instances, the Knight shift $K$ fails to track the temperature dependence of $\chi$ below a crossover temperature $T^* \sim 10-100$ K  (see Fig. \ref{fig:cecoin5shift}).  This Knight shift anomaly is ubiquitous among the heavy fermions, yet is not fully understood, and suggests \textit{a priori} that the hyperfine coupling itself may be temperature dependent.  This scenario would mean that all of the framework for understanding the Knight shift and \slrr\ in the superconducting state (and indeed the normal state) would not be robust, leaving one with little confidence that the NMR data can be interpreted in any meaningful way.

There are three theories of the Knight shift anomaly that have been proposed in the literature: two of these scenarios rely on a more local picture, and imply that the hyperfine coupling itself changes.  The first, postulated by Cox and coworkers  shortly after the discovery of the Knight shift anomaly in CeSn$_3$ \cite{CeSn3Knightshift}, assumes that the f-electron spins are screened at temperatures $T< T_K$, leading to a suppression of the bulk susceptibility, $\chi$ \cite{coxKSA}.  The Knight shift, on the other hand, measures the \textit{local} susceptibility and does not sense the screened moment.  Although this scenario could explain the CeSn$_3$ data, it fails to capture the behavior of the anomaly in several other heavy fermion systems where $K \nsim \chi$ below $T^*$.

Another scenario was originally postulated by Yasuoka and Fisk to explain the Knight shift data in CeCu$_2$Si$_2$, in which the hyperfine coupling between the ligand Cu and Si nuclei and the Ce 4f electron spin depended on the crystal field states of the $J=5/2$ multiplet \cite{YasuokaCeCu2Si2}.  As the temperature varies, the population of the CEF levels change and hence the relationship between $K$ and $\chi$ is modified.  In this case, the local hyperfine field at a particular nucleus will depend the particular electronic configuration on the coupled 4f atom, which fluctuates on a fast time scale.  The effect over the ensemble of nuclei would be a motional narrowing of the line, with an effective average field weighted by different hyperfine couplings to the susceptibilities of each of the CEF-split doublets. Data in \cecoin\ was originally analyzed in a similar fashion \cite{CurroAnomalousShift}, but was ruled out later because $T^*=50$ K in this compound does not correspond to any of the CEF splittings, but rather coincides with the coherence onset temperature where many of the other observed quantities, such as the resistivity, change character.

To date, the scenario that is most successful in capturing the behavior of a broad range of materials is the two-component hyperfine scenario, which asserts that the anomaly arises due to coherence in the Kondo lattice. In the Kondo lattice model, there are two sets of electron spins that contribute to the susceptibility: the conduction
electron spins, $S_c=1/2$, and the local moments of the f-electron, $J_f$.
Typically, the f-electrons are localized at high temperature, giving rise to strong Curie-Weiss behavior. However, these conduction and f-electrons experience several interactions which modify their behavior at low temperature.  The localized $f$-electron $J_f$ multiplet is split by a crystalline electric field (on the order of $\mathcal{H}_{\rm CEF}\sim$10 meV), so that for $T \ll \mathcal{H}_{\rm CEF}/k_B$ the magnetic behavior of this degree of freedom can be treated as a pseudospin doublet with an effective spin $S_f=1/2$ and a $g$ value that depends on the details of the crystal field parameters.  Secondly, the local moments and conduction electrons are coupled via a Kondo interaction
$\mathcal{H}_{\rm Kondo}\sim J\mathbf{S_c}\cdot\mathbf{S_f}$.  This interaction gives rise to two different effects: Kondo screening of the local moments, and exchange interactions (Rudemann-Kittel-Kasuya-Yosida, RKKY) between the local moments.  The Kondo interaction leads to a characteristic temperature $T_K=N(0)/k_Be^{-1/JN(0)}$, where $N(0)$ is the density of states at the Fermi level.  Below this temperature, the f spins are screened by the conduction electron spins, reducing their effective moment. On the other hand, the Kondo interaction also gives rise to an indirect exchange between local moments via polarization of the conduction electron medium. The characteristic temperature for the RKKY interaction is given by $T_{\rm RKKY}\sim (JN(0))^2$.  These two effects lead to two very different ground states: a spin liquid when the Kondo screening dominates, and long range magnetic order when the RKKY interaction dominates. Several years ago, Doniach considered the phase diagram of such a lattice of f-electrons, the Kondo lattice, and found that below a critical value of $JN(0)$, the system would undergo long range magnetic order of the f-electrons, whereas for greater values the system remains disordered \cite{doniach}.  This simple model describes some the essential physics, but fails to capture many details particularly in the vicinity of the quantum critical point (QCP), where superconductivity can emerge and the behavior of the normal state cannot be described as a Fermi liquid \cite{ColemanHFdeath}.  Despite many years of intensive theoretical work, there still remains no complete theory for the behavior of these heavy fermions in the vicinity of the QCP.

Recently, Pines and coworkers have taken the approach that the essential physics of the Kondo lattice can be elucidated by characterizing the various materials in terms of a common phenomenology \cite{NPF,YangDavidPRL,YangPinesNature}.  This two-fluid picture assumes that below a temperature $T^*$, the Kondo lattice is described by a coexistence of both local moment and heavy electron behavior.  $T^*$ is
a material dependent temperature on the order of 10 - 100 K.  This scenario offers an appealing microscopic picture of the  Knight shift anomaly. For $T>T^*$, $K$ is linearly
proportional to $\chi$. Since the dominant contribution to the
susceptibility is from the local moments, $S_f$, at these
temperatures, we conclude that the largest hyperfine field arises
from the local moments. In general, however, there can be also an
on-site hyperfine interaction to conduction electrons.  A more
complete description of the hyperfine interactions is given by:
\begin{equation}
\hat{\mathcal{H}}_{\rm hyp}=\gamma\hbar\mathbf{I}\cdot\left(
\mathbb{A}\cdot\hat{\mathbf{S}}_c+ \sum_{i\in
nn}\mathbb{{B}}\cdot\hat{\mathbf{S}}_f\right),
\end{equation}
where $\mathbb{A}$ is an on-site hyperfine tensor interaction to the
conduction electron spin, and $\mathbb{B}$ is a transferred
hyperfine tensor to the f spins \cite{CurroKSA}. Note that we consider here nuclear spins on the ligand sites, i.e., not on the f atom nucleus.   A
similar scenario is present in the high temperature superconductors
\cite{MilaRiceHamiltonian}. It is likely that the mechanism of the
on-site coupling $\mathbb{A}$ is probably due to a combination of
core polarization, and polarization of unfilled orbitals with s and
p symmetry.  The mechanism of the transferred interaction,
$\mathbb{B}$ may arise as a combination of orbital overlap
between the f electron states and the ligand atom wavefunctions,
and either a core polarization of the ligand atom or a dipolar interaction between the ligand nucleus and the polarized ligand atom p- and d- electrons.
For the remainder of this article, we assume that the hyperfine
parameters are material dependent constants, and will not discuss
their microscopic mechanism further.

Given the two spin species,
$\hat{S}_c$ and $\hat{S}_f$, there are three different
spin susceptibilities: $\chi_{cc}=\langle
\hat{S}_c\hat{S}_c\rangle$, $\chi_{cf}=\langle
\hat{S}_c\hat{S}_f\rangle$, and $\chi_{f\!f}=\langle
\hat{S}_f\hat{S}_f\rangle$.  The full expression for the Knight shift is
given by:
\begin{equation}
K(T)=A\chi_{cc}(T) +\left(A + \sum_{i\in nn}B_i\right)\chi_{cf}(T) +\sum_{i\in nn} B_i\chi_{ff}(T).
\end{equation}
where we have absorbed the g-factors
into the definition of the hyperfine constants
and dropped the tensor notation for notational simplicity \cite{CurroKSA}. The bulk susceptibility is given by:
\begin{equation}
\chi(T) = \chi_{cc}(T) + 2\chi_{cf}(T) +\chi_{ff}(T).
\end{equation}
Note that if $A=\sum_{i\in nn} B_i$, then $K\propto\chi$  for all temperatures. In general, this relation does not hold, and if $\chi_{cc}(T)$,
$\chi_{cc}(T)$, and $\chi_{cc}(T)$ have different temperature
dependences, then the Knight shift will not be proportional to
susceptibility, leading to a Knight shift anomaly at a temperature
$T^*$.

Since a complete solution of the Kondo lattice remains a challenge
for theory at present, there is no description of the temperature
dependence of $\chi_{cc}(T)$, $\chi_{cf}(T)$, and $\chi_{ff}(T)$.
However, using empirical observations Nakatsuji Pines and Fisk, and later Yang and Pines,
developed  a phenomenological two-fluid picture of the
susceptibility and specific heat in \cecoin\ which also works well
for understanding the Knight shift anomaly
\cite{NPF,CurroKSA,CurroPinesFiskMRS,YangDavidPRL}. In this picture, $\chi_{cf}(T)
\sim \left(1-T/T^*\right)\log(T^*/T)$ for $T<T*$, and $\chi_{cf}(T)
\sim 0$ for $T>T^*$, and we assume $\chi_{cc}$ is negligible for all
temperatures. This relation appears to hold for all
heavy fermion and mixed valent compounds measured to date down to the relevant
ordering temperatures; in fact, $\chi_{cf}$ scales with $T/T^*$ in all instances. For a detailed discussion we refer the reader to \cite{CurroKSA,curroSCES}. Recent
theoretical work  and DMFT calculations lend support for the two-fluid picture
\cite{SunKotliar,barzykin2fluid}, but a complete microscopic description is still lacking. Clearly, in the case of the heavy fermions it is the heavy quasiparticles that form the superconducting condensate, and the hyperfine coupling to these heavy quasiparticles is modified by the onset of coherence. Any experiment measuring the NMR properties in the superconducting state of a heavy fermion system must therefore somehow address the normal state Knight shift anomaly at some level.



\subsection{Superconducting order parameters}

\subsubsection{\cecusi}

\begin{figure}
\centering
\includegraphics[width=\textwidth]{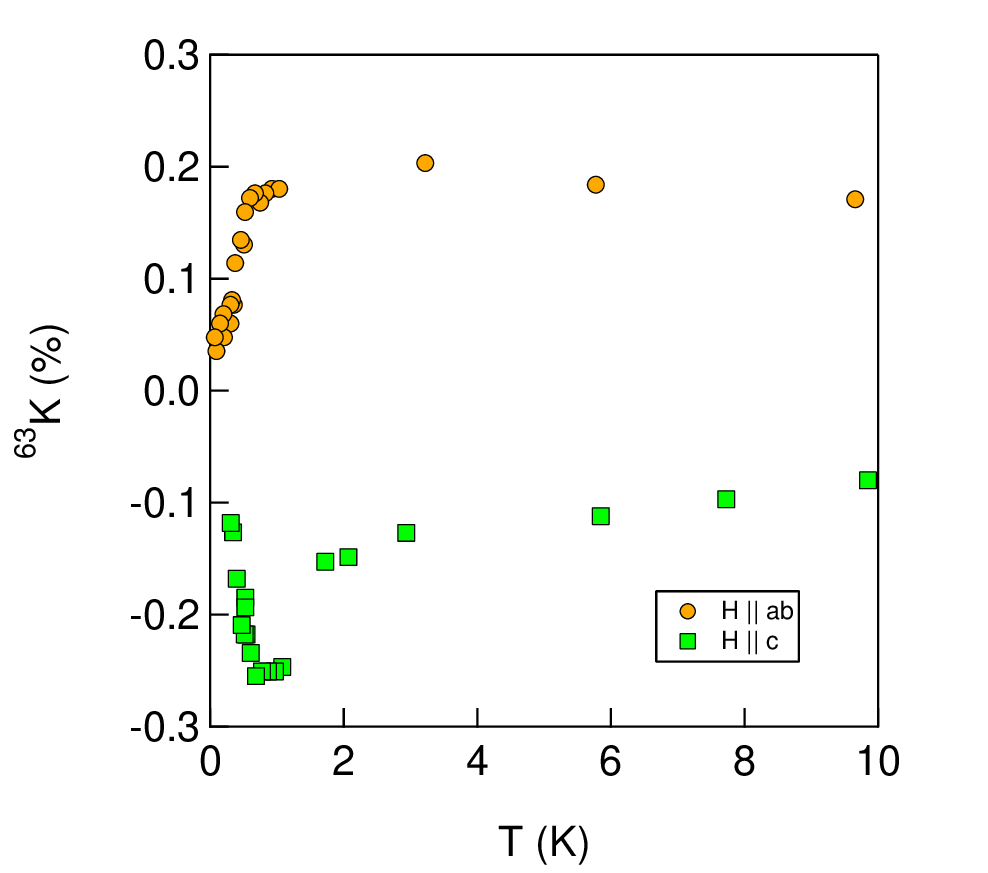}
\caption{The Cu Knight shift in \cecusi. The data are reproduced from \cite{CeCu2Si2KSC,YasuokaCeCu2Si2}.  The increase in $K_c$ below \tc\ is consistent with the behavior above \tc, similar to that observed in \cecoin. \label{fig:cecu2si2} }
\end{figure}

Since the discovery of the first heavy fermion superconductor, \cecusi\ in 1979, NMR has played a crucial role in determining the symmetry of the superconducting order parameter.  The Knight shift of the Cu in this system is suppressed below \tc, for fields oriented both parallel and perpendicular to the tetragonal c-axis (see Fig. \ref{fig:cecu2si2}) \cite{CeCu2Si2KSC}. Note, however, that for the field along the c-direction, the shift \textit{increases} below \tc.  This suggests that somehow the spin susceptibility increases below \tc, contrary to the behavior expected in a spin singlet superconductor.  In fact, in the normal state between \tc\ and $T^*\sim 170$ K, $K_c$ decreases whereas $\chi_c$ increases with decreasing temperature.  Naively, one might interpret this as a negative hyperfine coupling, so that below \tc\ when $\chi_{\rm spin}$ decreases (and vanishes in a spin singlet), the net effect on the total shift increases.  A more sophisticated interpretation in the two component model is that $(A-B)\chi_{cf}(T)$ is negative in the normal state, and that $|\chi_{cf}|$ decreases below \tc.  However, it is not straightforward to associate $\chi_{cf}$ with $\langle S_c S_f\rangle$ in this situation, as the character of the spins to which the nuclei are coupled in this heavy electron state are not simply described by $S_c$ and $S_f$, but rather in terms of quasiparticle excitations of a highly correlated electronic state.  Clearly, the susceptibility of these degrees of freedom is suppressed below \tc, but how the character of the local moments changes in the superconducting state is poorly understood. Spin lattice relaxation measurements in the superconducting state of \cecusi\ reveal a power law dependence; thus combined with evidence for spin-singlet pairing, it is likely that this material is a d-wave superconductor  \cite{cecu2si2T1}.

\subsubsection{UPt$_3$ - spin triplet superconductivity}

\begin{figure}
\centering
\includegraphics[width=4.0in]{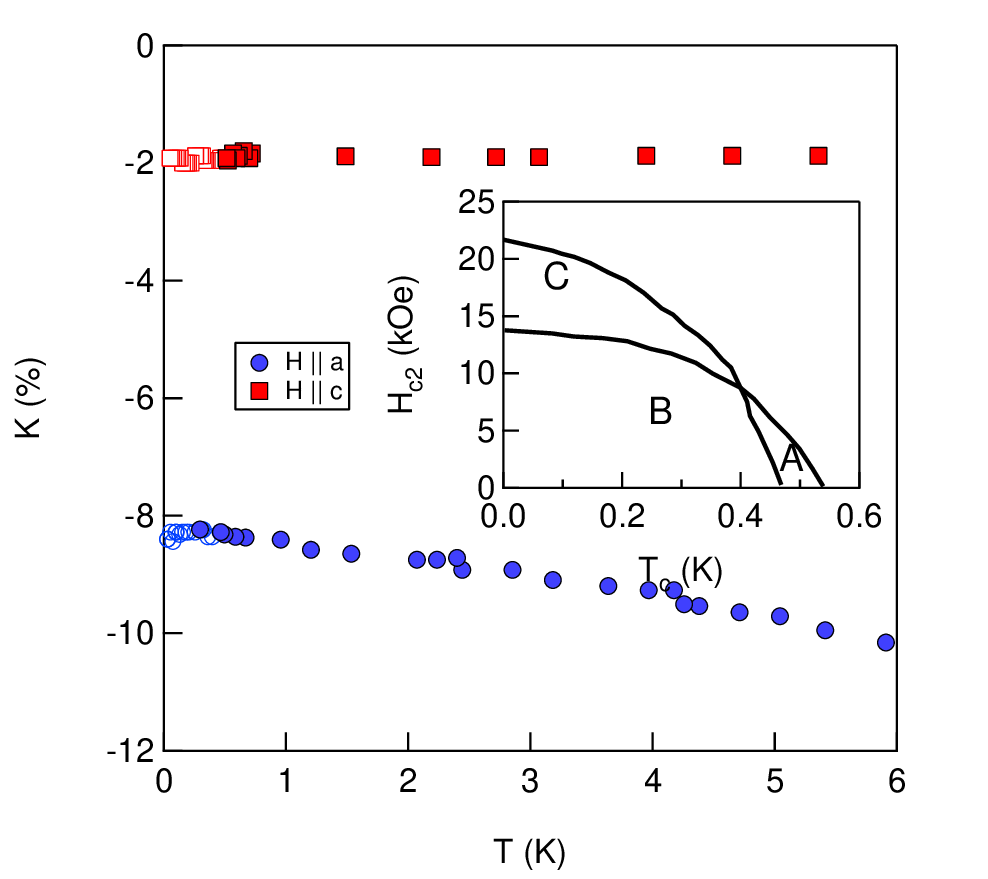}
\caption{The Knight shift of the Pt in UPt$_3$, reproduced from \cite{TouUPt3NMR}. Open points correspond to the superconducting state, whereas solid points are in the normal state. The Inset shows the $H-T$ superconducting phase diagram for field $H$ along the $c$ direction. \label{fig:upt3shift}}
\end{figure}

UPt$_3$ is a highly unusual superconductor that exhibits multiple phases within the $H-T$ phase diagram, that are most likely associated with different symmetries of the superconducting order parameter (see inset of Fig. \ref{fig:upt3shift}), and the majority of the data suggest that this is a spin-triplet superconductor \cite{flouquetUPt3,FlouquetUPt3gap}.  The multiple phases possible below \tc\ strongly suggest an order parameter with internal degrees of freedom. As described above, the response of the spin susceptibility below $T_c$ depends on the direction of $\mathbf{d}$ relative to the field $\mathbf{H}_0$.  In other words, the spin susceptibility of the superconducting condensate must be described by a tensor that may have off-diagonal components. In a single crystal of UPt$_3$, Knight shift measurements of the Pt clearly showed no change between the normal and superconducting states for both field orientations \cite{TouUPt3NMR}.  This result is clearly incompatible with spin singlet pairing, and furthermore suggests that the spin polarization of the condensate is not pinned to the lattice, as was found in Sr$_2$RuO$_4$ \cite{IshidaSr2RuO4}, but rather follows the direction of the applied field, maintaining the same anisotropy as the normal state.

\subsubsection{CeCoIn$_5$ and non-Fermi liquid behavior}

\cecoin\ is particularly interesting heavy fermion superconductor
 because it shows a rich behavior  of phenomena associated with the proximity of the system to an antiferromagnetic quantum critical point.  The superconducting state phase diagram of this material exhibits an unusual field-induced  transition to a new thermodynamic phase that lives only within a narrow regime between a field $H^*(T)$ and $H_{c2}$.  This field induced phase will be discussed in detail below.  For fields $H < H^*(T)$, $K(T)$ and \slrr\ behave in a similar manner to those in \cecusi.  The Knight shift of the In(1), In(2) and Co all change below \tc, but the sign of the change varies from site to site \cite{CurroAnomalousShift,kohori115shift}. The In(1) shifts down below \tc\ in both orientations, the Co and In(2) shift up for fields along $c$ and down for fields along $ab$, and the In(2)$_{\perp}$ shifts down for fields along $ab$ (the In(2) splits into two distinct sites, In(2)$_{||}$ and In(2)$_{\perp}$ for fields along $ab$ because it is in a low symmetry site).
Empirically, 
if $K(T)$ is increasing above \tc, then it decreases below, and if $K(T)$ is decreasing above \tc, then it increases above \tc.  As in the \cecusi\ case, this behavior can be understood in terms of a negative hyperfine coupling, and hence the spin susceptibility decreases below \tc.

The detailed behavior of the Knight shift as a function of field in the superconducting state of this material shows a dramatic increase with increasing field \cite{mitrovicKvsHCo115}.  $\chi_{s}(T=0,H)$ should vary linearly with field for a d-wave superconductor because the Zeeman shift of the quasiparticle dispersion leads to a linear increase of the density of states at $E_F$ \cite{saulsKvsH}.  In other words, the point nodes develop into arcs in $\mathbf{k}$-space, as in the case of the Doppler shift.   $K(T=0,H)$ of the In(1) shows a gradual increase, and eventually gets quite steep as $H$ approaches $H^*(0)$ and $H_{c2}$ in this material.  This behavior strongly suggests d-wave pairing, with line nodes.

\begin{figure}
\centering
\includegraphics[width=4.0in]{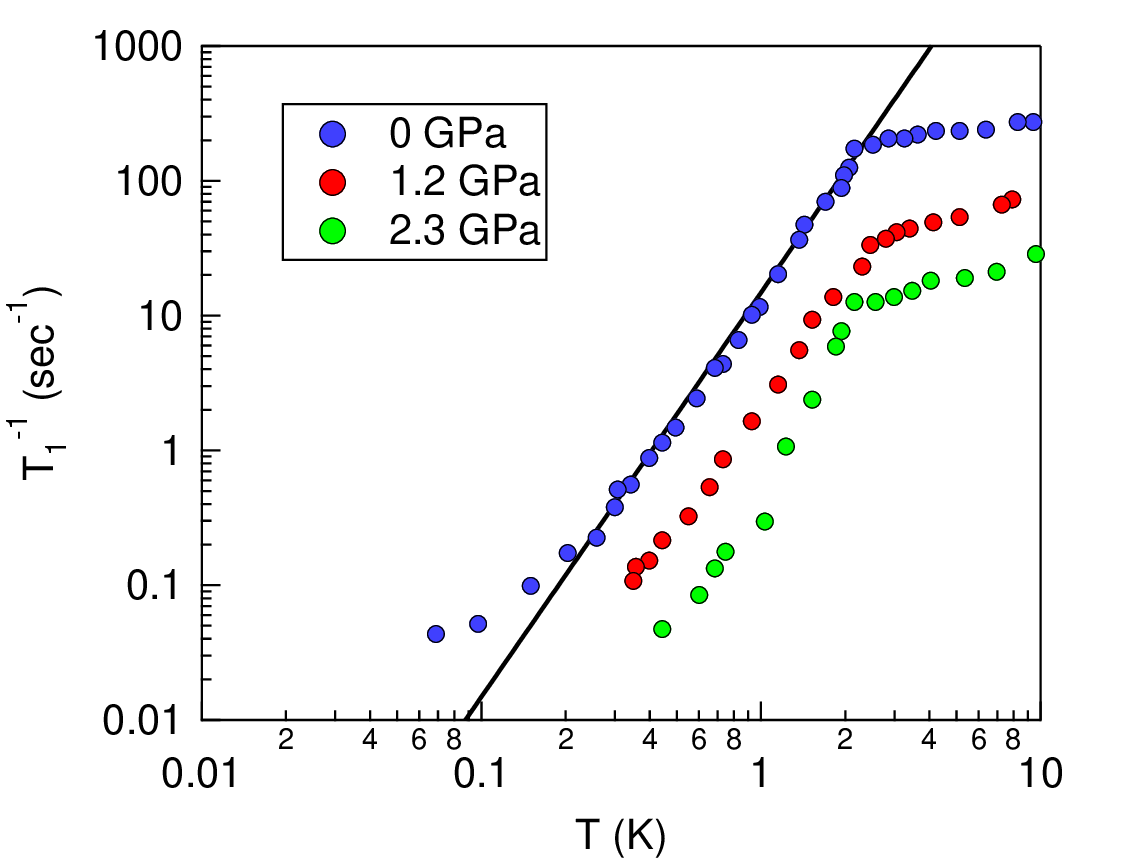}
\caption{\slrr\ versus temperature in \cecoin\ measured by NQR at zero applied field, reproduced from \cite{kohoriCeCoIn5pressure}. The solid black line is a fit to $T^3$, as expected for line nodes in the superconducting gap. \label{fig:CeCoIn5T1SC} }
\end{figure}

Measurements of \slrr\ are straightforward in the \cecoin\ because the In is spin $I=\frac{9}{2}$ and has a large quadrupolar frequency.  The data of several groups show consistent $T^3$ behavior, consistent with line nodes, as seen in Fig. \ref{fig:CeCoIn5T1SC}.  Measurements of \slrr\ in field have not been performed, but may be important since recent small angle neutron scattering data suggest the presence of antiferromagnetic vortex cores in this material, which would contribute extra relaxation channels \cite{BianchiSANS}.  This antiferromagnetism is probably related to the proximity to a quantum critical point.

\begin{figure}
\centering
\includegraphics[width=4.0in]{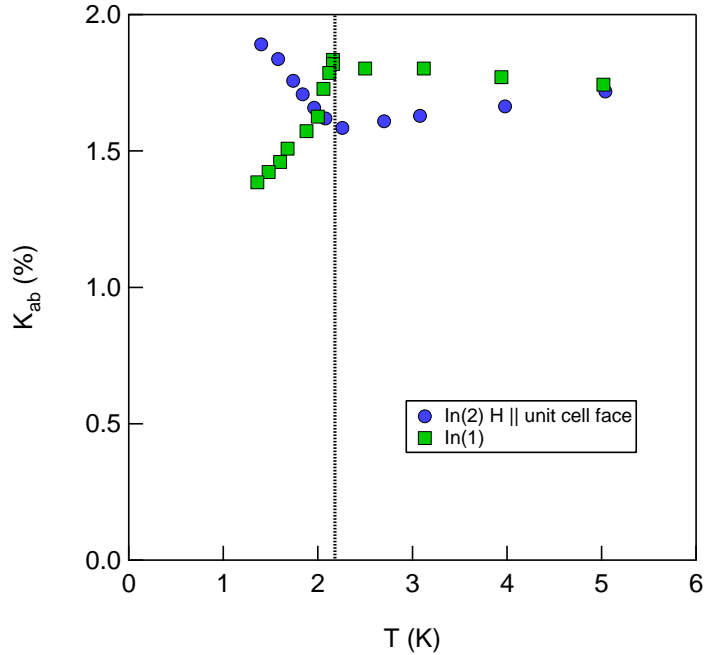}
\caption{The Knight shift of the In(1) and the In(2) in \cecoin, reproduced from \cite{CurroAnomalousShift}. Note that the shift of the In(1) \textit{increases} below $T_c$, similar to that of the Cu in \cecusi. \label{fig:cecoin5shiftSC} }
\end{figure}

\subsubsection{PuCoGa$_5$ and the bridge to high \tc\ superconductivity}

\begin{figure}
\centering
\includegraphics[width=4.0in]{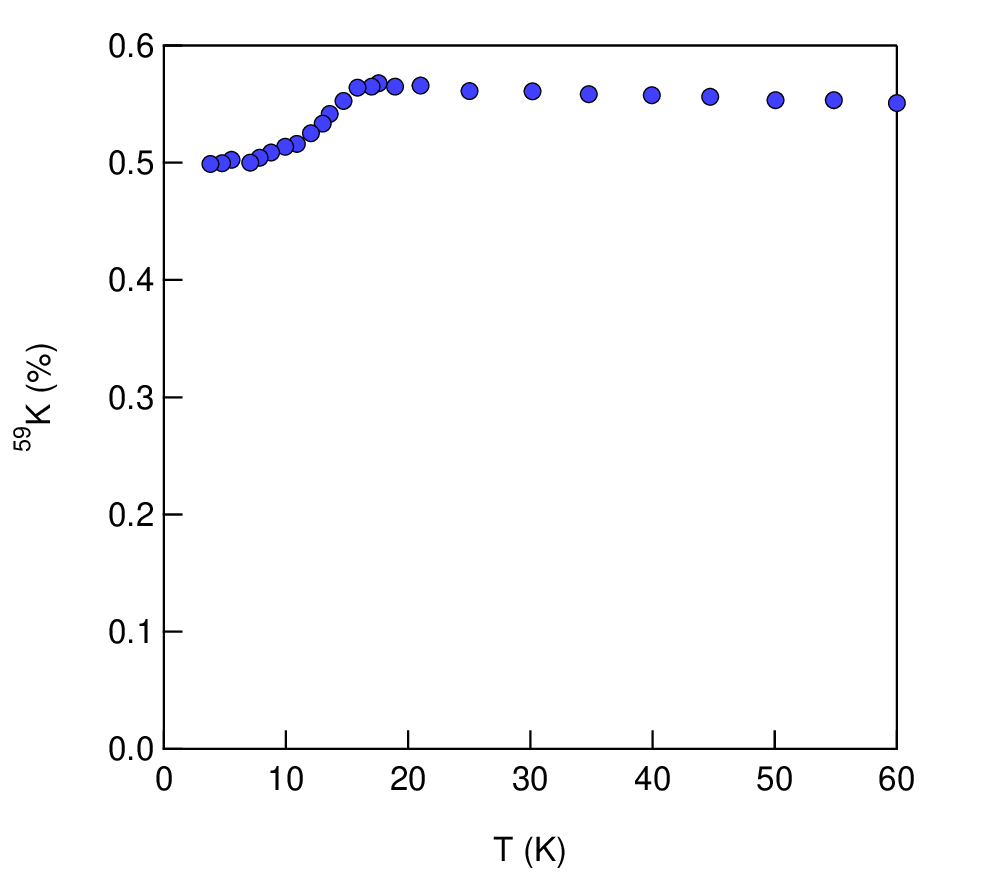}
\caption{The Knight shift of the Co in \pucoga, showing the suppression of the spin susceptibility below $T_c$, indicative of spin singlet pairing. \label{fig:pucoga5shift} }
\end{figure}

Until the year 2002, the class of unconventional superconductors seemed to consist of three separate subclasses: the \hightc\ cuprates, the organic superconductors, and the heavy fermion superconductors.  The cuprates, with transition temperatures around 100-150 K, and the organic superconductors, with transition temperatures on the order of 5-10K, have structures based on two dimensional doped Mott insulators.   The heavy fermions have transition temperatures of 1-2 K, are good metallic systems with either Ce or U atoms in the lattice.  Since the physics of the basis materials seem to be quite different, it is natural to suspect the nature and origin of the superconductivity may be different as well.  However, these different sets of materials have similar phase diagrams, with superconductivity emerging near to antiferromagnetism, and in most cases the superconductivity appears to be unconventional.  In 2002, \pucoga\ was found to be superconducting at 18.5 K.  This result was surprising because the transition temperature is so high and the structure is identical to that of \cecoin.  By going from the 4f to the 5f series, the transition temperature increased by an order of magnitude.  One of the leading questions was whether the superconductivity in \pucoga\ is conventional or unconventional.  NMR played a crucial role, as one of the only techniques available to deal with the radioactive plutonium compound.  Knight shift measurements of the Co and Ga clearly revealed a spin singlet superconductor, and \slrr\ measurements showed a nodal gap. The fact that \pucoga\ was a heavy electron unconventional superconductor suggested then, that the gap between the heavy fermions and the \hightc\ cuprates may be artificial, and that unconventional superconductivity can arise over a continuous range of temperatures, possibly controlled by the scale of antiferromagnetic exchange in the material \cite{CurroPuCoGa5,uedareview}.  Another Pu-based superconductor, \purhga\ with $T_c \approx 9$ K, was discovered shortly after \pucoga, and NMR measurements of $K$ and \slrr\ also reveal an nodal, spin-singlet superconductor \cite{walstedtpurhga5}.  Quite recently in 2007, another actinide-based superconductor was discovered, NpPd$_5$Al$_2$ ($T_c=4.9$ K) \cite{NpPd5Al2discovery,griveauNpPd5Al2}. NMR measurements reveal line nodes and spin-singlet behavior in this system, also suggestive of d-wave superconductivity \cite{NpPd5Al2NMR}.  Clearly, the actinide series holds many surprises to come, and NMR will play a critical role in determining the properties of any superconducting materials.

There are other heavy fermion superconductors, in particular UBe$_{13}$, UPd$_2$Al$_3$, UNi$_2$Al$_3$, where NMR measurements in the superconducting state have shown power-law behavior of the \slrrtext\ in the superconducting state, but the Knight shift measurements and overall interpretation have remained inconclusive.  We refer the reader to \cite{toukitaokareviewT1} for more details. Other materials, such as \cerhin\ and CeIn$_3$ become superconducting under pressure. In these cases, \slrr\ measured by NQR reveals $T^3$ behavior, but depending on the pressure there is evidence for a $T-$ linear term at lower temperatures \cite{kitaokacecoexistencereview,kitaokaCeRhIn5pressureGapless,kitaokaCeIn3}.  This crossover behavior at low temperature is poorly understood, but may be related to the coexistence of antiferromagnetism \cite{ZhengCeRhIrIn5}.

\subsection{The field-induced phase in \cecoin}

\begin{figure}
\centering
\includegraphics[width=\textwidth]{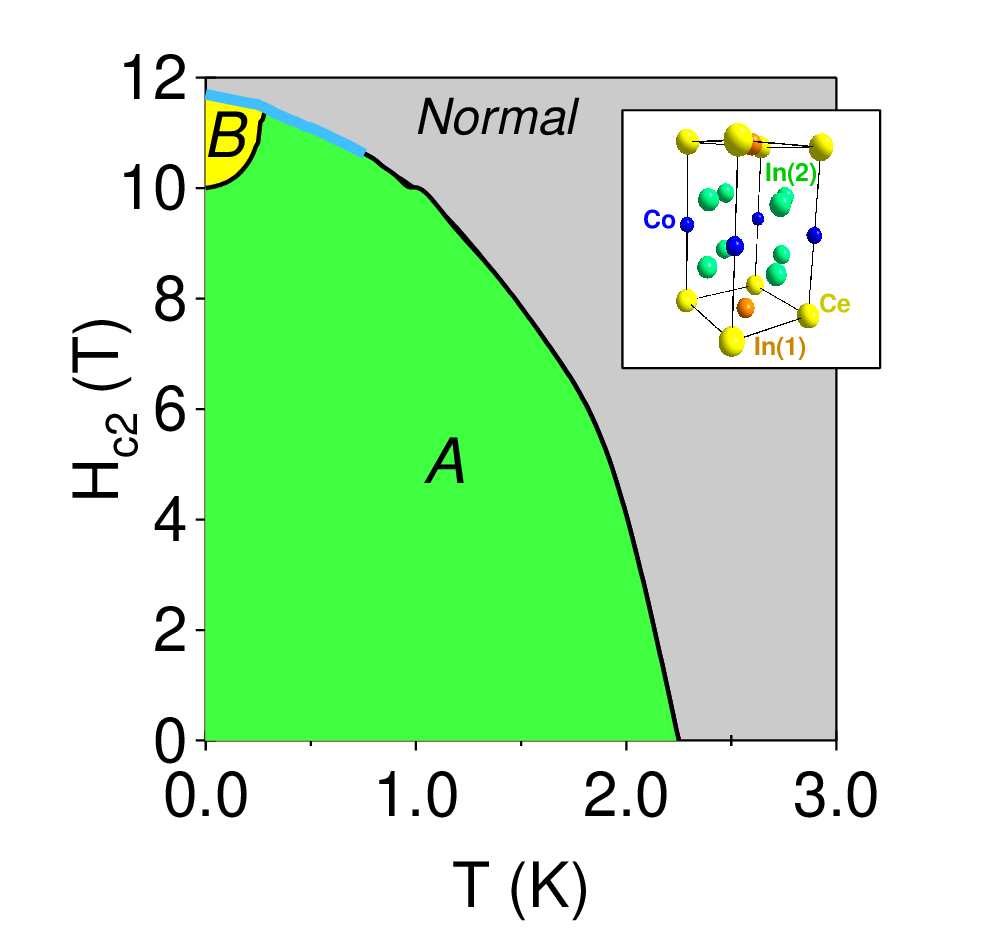}
\caption{The superconducting phase diagram of \cecoin\ (structure
shown as inset). The solid black lines represent second order phase
transitions and the solid blue line represents a first order
transition.  The A phase is a conventional Abrikosov vortex lattice, and the B phase is the field-induced phase that may be related to an FFLO phase. The phase boundary between the A and B phases is referred to in the text as either $T^*(H)$ or $H^*(T)$.\label{fig:cecoinphasediagram} }
\end{figure}

Some of the most striking and unusual behavior exhibited by \cecoin\ emerges at ambient pressure in the superconducting state. In 2002, heat capacity measurements close to $H_{c2}$ revealed an unexpected first order transition between the normal state and the superconducting state, and further measurements indicated the presence of a second order phase transition within the superconducting state a temperature $T^*(H)$ \cite{andrea,romanFFLO,radovanCeCoIn5FFLO}.  This phase was initially identified as a possible Fulde-Ferrell-Larkin-Ovchinnikov (FFLO) phase, which was postulated to exist in superconductors with exceptionally large spin susceptibility \cite{ff,lo,maki}.  Magnetic fields can suppress superconductivity by two mechanisms: orbital limiting or Pauli limiting.  In an orbital limited superconductor, the highest magnetic field that the system can support is limited by the critical current density, $J_c$.  The London equations show that superconductors respond to external fields by creating supercurrents to screen the internal field, but if the supercurrent exceeds $J_c$ then the system can no longer support superconductivity \cite{tinkham}.  In a Pauli-limited superconductor, the maximum magnetic field is determined by the spin susceptibility of the quasiparticles.  For sufficiently large fields, the Zeeman energy of the quasiparticles can exceed the superconducting condensation energy, in which case the material can no longer support superconductivity.  Typically the Pauli susceptibility of superconducting materials is small enough that the limiting field is determined by the orbital limit. However, since heavy fermions typically have large susceptibilities due to the enhanced effective mass, the Pauli limit may be reached in principle.

In Pauli limited superconductors, the transition between the normal and superconducting phases can become first-order, and in some cases a new phase can emerge, in which the superfluid density becomes modulated over large length scales.  In the FFLO phase, the order parameter $\Psi \sim \Psi_0 \exp(-i\mathbf{q}\cdot\mathbf{r})$ or $\Psi \sim \Psi_0 \cos(\mathbf{q}\cdot\mathbf{r})$, where $\mathbf{q}=\mathbf{k}_{F,\uparrow} - \mathbf{k}_{F,\downarrow}$, where $\mathbf{k}_{F,\uparrow,\downarrow}$ are the Fermi wavevectors of the spin-up and spin-down Fermi surfaces.  The wavelength, $\Lambda = 2\pi/q$, is therefore several times the unit cell length, $a$.  The minimum possible distance must be $\Lambda_{min} = \xi$, the superconducting coherence length, since the condensate cannot respond on a length scale shorter than $\xi$.  In this phase, the wavefunction vanishes at nodes in real space,  and the unpaired up-spin quasiparticles reside in these nodes.  This phase was predicted to exist nearly forty years ago, but experimental evidence for its existence has been sparse.  Likely candidates include the organic superconductor (TMTSF)$_2$ClO$_4$ and \cecoin\ \cite{organicFFLO,radovanCeCoIn5FFLO}.  The strongest evidence for the existence of this phase in \cecoin\ stems from the similarity of the phase diagram, shown in Fig. \ref{fig:cecoinphasediagram}, to calculations.  All of the theoretical calculations, however, are based on either a non-interacting or weakly interacting Fermi liquid \cite{grafknight,graf}.  The \cecoin, however, clearly has more physics at play than a simple Fermi liquid, as there are strong f-electron exchange interactions (RKKY) and Kondo screening of the moments, leading to non-Fermi liquid behavior in the normal state \cite{romanQCPCeCoIn5,taillefair}.    Therefore, it is not obvious \textit{a priori} that this new phase can be described by the simple theories of FFLO developed for Fermi liquids. Indeed, NMR evidence suggests that the situation is more complex. We therefore refer to this phase simply as the B phase, in order to distinguish it from the classical FFLO phase described by these theories.

\begin{figure}
\centering
\includegraphics[width=\textwidth]{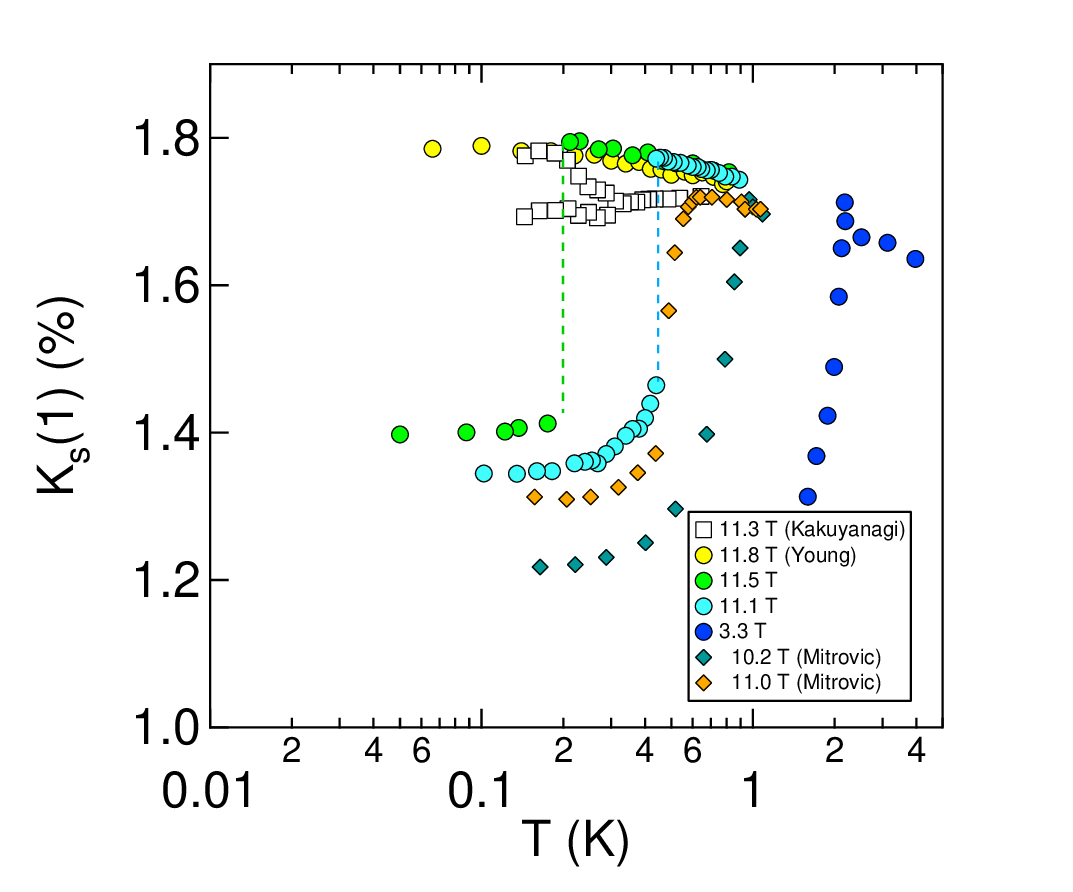}
\caption{The Knight shift of the In(1) in \cecoin\ for $H || ab$ as measured at various field and reported by Kakuyanagi et al. \cite{kumagaiCeCoIn5}, Young et al. \cite{CurroCeCoIn5FFLO}, and  Mitrovic et al. \cite{mitrovic}. Dotted lines are guides to the eye for discontinuous transitions. \label{fig:FFLOcomparisonshift} }
\end{figure}

The first NMR experiments in the B phase were reported by \cite{kumagaiCeCoIn5}.  The spectra of the In(1) in this report showed dramatic changes as the temperature was lowered through  $T^*$.   Below $T^*$, the spectra revealed two components, one peak at slightly reduced frequency, and a second smaller peak at the resonance frequency of the normal state (see Fig \ref{fig:FFLOcomparisonshift}).  These data were interpreted as evidence for coexistence of both superconducting and normal regions, consistent with the FFLO picture of normal planes sandwiched between superconducting layers.  One puzzling aspect of this data, however, is that the spectra seemed to reveal a continuous evolution rather than a first order transition.  Secondly, the change $\Delta K = K(T_c) - K(T=0)$ of the Knight shift of the lower (superconducting) peak was much smaller than $\Delta K$ reported previously at lower fields \cite{CurroAnomalousShift,kohori115shift}.  Subsequent measurements by other groups at high fields disagreed with this initial report, both quantitatively and in interpretation of the data \cite{mitrovic,CurroCeCoIn5FFLO}.  The data sets are compared in Fig. \ref{fig:FFLOcomparisonshift}.   Mitrovic et al. argued that the spectra not only showed much larger $\Delta K$, but also that the intensity of the signal is suppressed by roughly an order of magnitude below \tc.  The original Kakuyanagi et al. data showed no decrease in intensity below \tc, although in practice the signal is suppressed because the rf penetration is reduced.  Young et al. reported discontinuous jumps in the Knight shift, with significantly larger $\Delta K$ values.  As seen in Fig. \ref{fig:FFLOcomparisonshift}, the Young and Mitrovic data agree well with one another, but not with that of Kakuyanagi.  Recently, Mitrovic et al. published a comment on the original Kakuyanagi paper, in which they argue that the Kakuyanagi data are incorrect and are the extrinsic effects of rf heating of the sample \cite{mitroviccomment}.  Mitrovic et al. were able to reproduce the two peaks observed in the Kakuyanagi data by systematically increasing the rf power.  Since the transition is first order, it is reasonable to expect that there may be some coexistence of both normal and superconducting regions near \tc, and for excessive rf heating (Joule-heating by induced currents in the metallic sample) this coexistence may persist over much larger temperature ranges.

\begin{figure}
\centering
\includegraphics[width=\textwidth]{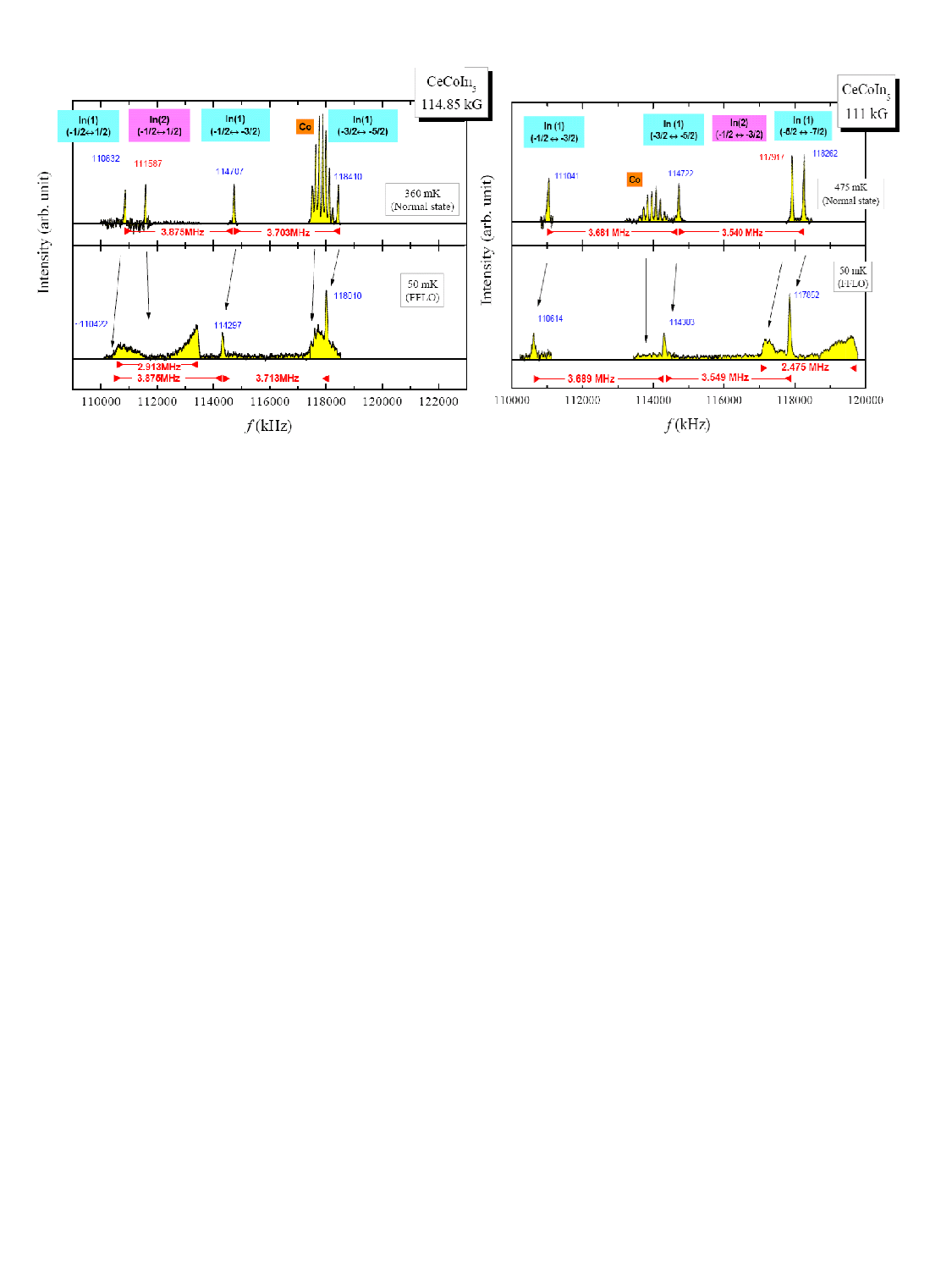}
\caption{(Left panel) NMR spectra in the normal state ($T =$ 0.475 K) and in the B phase ($T =$ 0.050 K) of \cecoin\ at 11.1 T, showing the spectra of the Co ($I=7/2$), several of the  In(1) ($I=9/2$)  transitions, and one of the In(2) transitions. Clearly, several differences emerge in the B phase, most notably the broad double-peak form of the In(2), which is characteristic of an incommensurate magnetic order.  In contrast, the In(1) and Co show little change. (Right panel) Similar spectra, but at a higher field of 11.485 T, in the normal state ($T=0.36$ K) and in the B phase ($T=0.05$ K).  \label{fig:cecoinBphasespectra} }
\end{figure}

Young et al. also  report data on the Co and In(2) sites over a series of temperatures crossing from the normal state, through the mixed phase, and into the field-induced phase.  They found that the Co, the In(1) and the In(2) resonances behaved very differently from one another  in both the mixed phase and the B phase. As seen in Fig. \ref{fig:cecoinBphasespectra}, the In(2) reveals a broad two-peak spectrum consistent with antiferromagnetic order.  The resonances of the other two sites, on the other hand, remain relatively sharp.  Young et al. argued that since the response of the three sites in the unit cell could not be explained by the conventional picture of the FFLO phase, in which the inhomogeneity is a long-range phenomenon with $\Lambda \gg a$, that a more complex scenario must be present.  In particular, the B phase exhibits long range antiferromagnetism.  The spectrum of the In(2) in this phase is similar to that of the In(2) in the antiferromagnetic state of \cerhin, and \cecoincdx\ \cite{curromagnetism115s}.  Assuming isotropic hyperfine couplings, the NMR data can be explained by an incommensurate structure with $\mathbf{Q}=(\frac{1}{2} + \delta, \frac{1}{2}, \frac{1}{2})$ and moments along the applied field direction, which gives rise to a finite hyperfine field at the In(2) sites, but not at the In(1) or Co sites, as observed in the data.  This wavevector represents only one possible magnetic structure that can explain the set of hyperfine fields observed in the B phase and is not unique, but is the most likely one if the hyperfine couplings to the In(2) are purely isotropic.


Recently elastic neutron scattering experiments were reported in the B phase of \cecoin, which clearly reveal an incommensurate magnetic order and thus confirming the work of Young and coworkers \cite{KenzelmannCeCoIn5Qphase}.  In these experiments, the field was aligned along [1-10], and the Ce moments were observed to lie along [001] with a magnitude of 0.15$\mu_B$, and wavevector $\mathbf{Q}=(\frac{1}{2} + \delta, \frac{1}{2}+\delta,\frac{1}{2})$ with $\delta=0.06$. In this case, the incommensurate modulation lies in the [110] direction, perpendicular to both the applied field and the ordered moments.  These observations contrast somewhat with the original NMR measurements with field along [100].  Although it is possible that the ordering wavevector and the direction of the moments may change when the applied field is rotated by 45$^{\circ}$, the discrepancy between the two measurements probably arises from insufficient information about the hyperfine tensor of the In(2).  In the NMR report, the tensor was assumed to be isotropic, but if this constraint is relaxed, then other magnetic structures consistent with the spectra of the In(2) emerge as possible candidates. In particular, if the tensor has dipolar symmetry, then ordered moments along [001] can give rise to the observed spectra.

One may then ask whether the B phase is an FFLO phase or an antiferromagnetic one.  In fact, identifying this phase as an FFLO phase is nearly impossible with NMR, since one would have to detect a change of phase of the order parameter along the spatial modulation direction.  However, bulk measurements of the phase diagram are consistent with theories for an FFLO phase in a Pauli limited superconductor \cite{graf,grafknight,maki,dwaveenergy2}.  Furthermore, NMR measurements clearly show no sign of any magnetism for fields $H > H_{c2}$.  If the B phase were a secondary phase that either competes with, or coexists with the superconductivity, then it should persist above $H_{c2}$ in the normal state.  Indeed, in \cerhin\ under pressure, there is a field-induced phase that coexists with superconductivity below $H_{c2}$ and exists on its own above (see Fig.) \cite{tuson}.  In contrast, the B phase of \cecoin\ exists only within the superconducting phase.  This might be explained by a subtle change in the delicate balance between the RKKY interaction and Kondo screening of the Ce moments.  Indeed, the onset of a superconducting gap can modify both these interactions, and may have a stronger effect on the Kondo screening \cite{FradkinKondoSC,wrobel}.  Therefore, for $H \sim H_{c2}$, where the superconducting condensation energy, the RKKY interaction and the Kondo interaction are all of the same magnitude, it might be possible that the Kondo interaction is suppressed in the presence of the superconducting gap, and the balance is tipped in favor of the magnetic order. Above $H_{c2}$, when the gap vanishes, the balance is reversed, and magnetism would disappear.

An alternative explanation is that the driving element for the B phase is truly an instability towards the FFLO state, and the the antiferromagnetism grows in as a secondary parameter that may nucleate in the normal planes of the FFLO phase.  This scenario offers an explanation for the pressure dependence of the superconducting phase diagram, which shows that with increasing pressure the volume of phase space occupied by the B phase increases \cite{nicklasPHTdiagramCeCoIn5}.  Pressure tunes the \cecoin\ away from the antiferromagnetic quantum critical point, and suppresses the antiferromagnetic fluctuations \cite{sidorov,kohoriCeCoIn5pressure}.  If the B phase were driven by an antiferromagnetic instability, one would expect the opposite trend with pressure.
 In fact, there is evidence even within the mixed phase for antiferromagnetism within the vortex cores of the Abrikosov vortex lattice \cite{CurroCeCoIn5FFLO,BianchiSANS}.  Small angle neutron scattering (SANS) data for the field along the c-direction indicate that the form factor for the vortex lattice increases with field, rather than decrease as expected for a conventional vortex lattice \cite{brandt}.  Furthermore, the NMR linewidths of the Co, In(1) and In(2) in the mixed phase are significantly larger than expected by Eq. (\ref{eqn:VLmomenthighfield}).  In fact, the second moment depends on the particular nucleus, which cannot be explained if the broadening is purely due to orbital magnetisation.  Rather, there must be a spin magnetization contribution as well, which will give rise to different hyperfine field distributions at the different nuclei.  This inhomogeneous spin distribution is probably due to local moments in the vortex cores.  Bulk magnetization measurements also indicate a strong paramagnetic contribution in the mixed phase, which is further evidence for magnetism in the cores \cite{tayama}.  A secondary antiferromagnetic order parameter that nucleates in the vortex cores where the superconducting order parameter goes to zero is a natural expectation of a Ginsburg-Landau description of competing order parameters, and is also observed in the \hightc\ cuprates \cite{demler,lakescience}.  If the superconducting order vanishes at nodal planes, rather than simple vortex lines, then it may be easier for a competing order parameter to establish long range order.


 Importantly, the neutron scattering data of Kenzelmann et al. reveal that the ordering wavevector $\mathbf{Q}$ is \textit{independent of applied field} in the B phase.  This result is inconsistent with an FFLO scenario, where the instability is driven by the Pauli susceptibility.  On the other hand, using a Landau phenomenological theory for the coupling between the field and the superconducting order parameter, Kenzelmann et al. argue that the most likely coupling term leads to a superconducting gap function that carries finite momentum.  It appears, then, that this phase consists of a pairing of electrons with different wavevectors, as in an FFLO phase, but the driving interaction may be an antiferromagnetic instability instead of the large Zeeman term that leads to a conventional FFLO picture.  Clearly, this new phase offers a wealth of new physics, and NMR is likely to play a central role in its study.

\subsection{Probing the coexistence of antiferromagnetism and superconductivity}

\begin{figure}
\centering
\includegraphics[width=\textwidth]{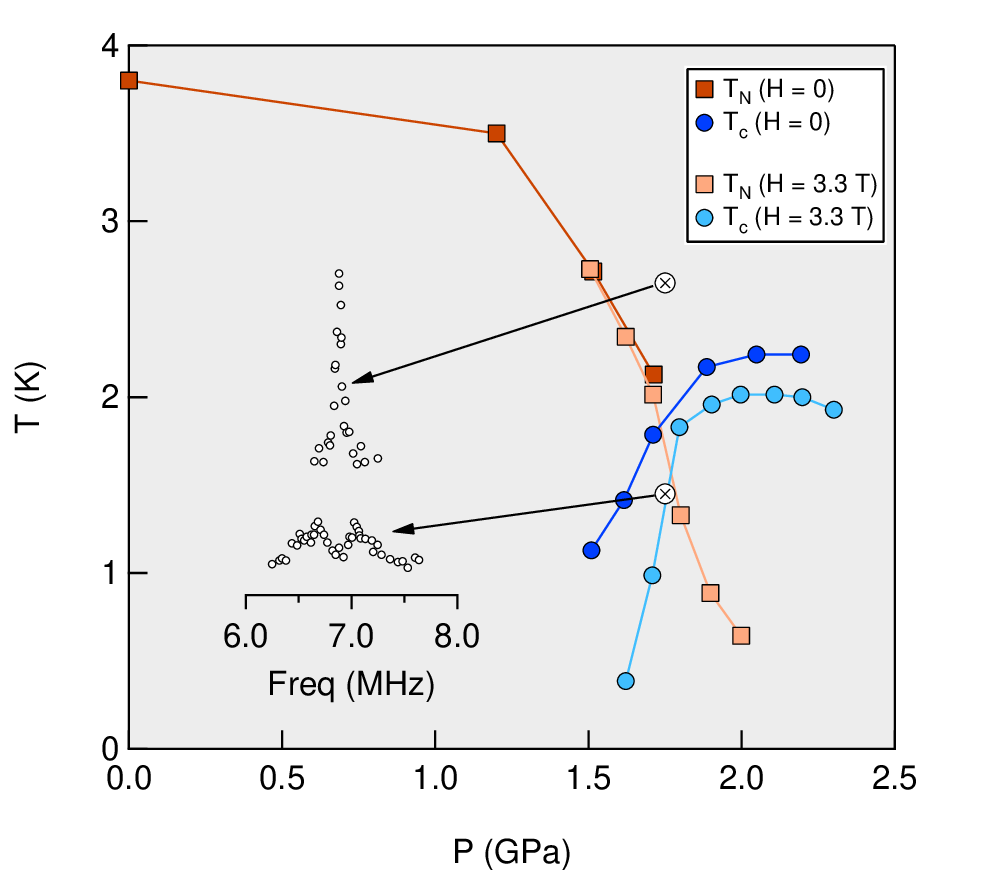}
\caption{The temperature-pressure phase diagram of \cerhin\ at zero field and at 3.3 T (reproduced from \cite{tuson}).  The spectra of the In(1) $(\pm1/2\leftrightarrow\pm3/2)$ NQR transitions (from \cite{kitaokaCeRhIn5coexistence}) are shown at the indicated points in phase space, at $H=0$. In the range between 1.5 - 1.8 GPa, antiferromagnetism and superconductivity coexist microscopically.  In field, the region of coexistence extends up to higher pressure up to 2.1 GPa (not shown) where deHaas-van Alphen experiments indicate a change of the Fermi surface \cite{onukiCeRhIn5dHvA}. \label{fig:Rh115phasediagram} }
\end{figure}

 NMR has played a central role in studies of coexisting antiferromagnetism and superconductivity in a number of heavy fermion compounds.  These two order parameters may arise in different channels, for example if there are more than one type of magnetic atom in the unit cell \cite{borocarbideMandSCneutrons}.  This may be relevant for several U-based heavy fermions, including UPd$_2$Al$_3$, UPt$_3$, URu$_2$Si$_2$, and UNi$_2$Al$_3$, since the U atom in these cases is typically has more than one 5f electron.  Here we focus on studies of coexisting orders for Ce compounds, in which there is only one channel available for the 4f$^1$ electronic degree of freedom from the Ce$^{3+}$ ions.  The same degree of freedom is responsible for the local moment magnetism, the heavy fermion state, and the unconventional superconductivity.  This is clearly seen in NMR studies of \cerhin\ under pressure \cite{kitaokaCeRhIn5pressureGapless,kitaokacecoexistencereview,kitaokaCeRhIn5coexistence}.  The temperature-pressure phase diagram of \cerhin\ is shown in Fig. (\ref{fig:Rh115phasediagram}).  For pressures between $\sim 1.5-1.8$ GPa, and $T<T_c$ the system shows bulk superconductivity microscopically coexisting with antiferromagnetism, as seen in the NQR spectra of the In(1).  The spectra clearly show that all of the In(1) nuclei in the same experience a static hyperfine field (see \S2.5).  If the material consisted of an inhomogeneous coexistence of regions that are superconducting but not magnetic, and regions that are antiferromagnetic but not superconducting, then the spectra would show two sets of resonances: one that is split by the internal field corresponding to the antiferromagnetic regions, and one that is not split, corresponding to the superconducting regions.  In other words, the spectra would be a combination of both the normal state spectrum and the ordered state spectrum as shown in Fig. (\ref{fig:Rh115phasediagram}).  Clearly, this is not the case.  Furthermore, \slrr\ measurements clearly show anomalies at both $T_N$ and \tc.  Although this observation is not as strong an argument for coexistence as the spectral evidence, it does support the interpretation since presumably the same set of nuclei experience both phase transitions. It should be pointed out, however, that \slrr\ is measured by fitting the relaxation of the bulk magnetization signal, which can in principle be composed of two signals: $M(t) = M_1(t) + M_2(t)$, where each set experiences only one of the two phase transitions.  In this case, the magnetization recovery will show multiple relaxation time scales.  Therefore, it is crucial to verify whether the magnetization recovery shows a single or multiple relaxation time scales \cite{ricardomultipleT1}.

In contrast to the \cerhin\ under pressure, NQR experiments in CeIn$_3$ reveal antiferromagnetism and superconductivity do not coexist microscopically, but are phase separated \cite{kitaokaCeIn3}.  Indeed, the In NQR spectrum consists of a superposition of the static broad line associated with the antiferromagentic order, and a narrow line associated with paramagnetism (the superconducting regions).  The relative volume fractions of these two regions vary as a function of pressure, with antiferromagnetism giving way to superconductivity, and the maximum in \tc\ occurs when the two volume fractions are the same. In both the \cerhin\ and the CeIn$_3$ case, the entropy associated with the two phase transitions arises entirely from single Ce 4f electron.  It is not clear why this degree of freedom can exhibit multiple orders simultaneously in the case of the \cerhin, and not in the CeIn$_3$, but may be related to the presence of a first order phase transition in the latter case \cite{kitaokaCeIn3}.  As a third example, in CeCu$_2$Si$_2$, superconductivity and antiferromagnetism coexist microscopically in some portions of the phase diagram, and in other regions the antiferromagnetism abruptly disappears when superconductivity sets in \cite{kitaokacecoexistencereview,stockertNS,stockertMSR}.  In the latter case, a possible scenario is that the antiferromagnetism is a spin density wave (SDW) order, and both order parameters compete for regions of the Fermi surface.  In contrast, both the fluctuations and the phase diagram in \cerhin\ suggest that a Fermi surface competition is not the case, but rather a dramatic Fermi surface reconstruction occurs in which the Ce 4f electrons change from localized to itinerant character \cite{onukiCeRhIn5dHvA}.

It is interesting to note that in states with coexisting antiferromagnetism and superconductivity \slrr\ typically exhibits a crossover in behavior in the superconducting state.  This is seen in \cerhin\ under pressure \cite{kitaokaCeRhIn5pressureGapless}, in CeRh$_{1-x}$Ir$_x$In$_5$ \cite{ZhengCeRhIrIn5}, and in \cecoincdx\ \cite{ricardomultipleT1,CeCoIn5CdDroplets}.  In \cerhin\ and CeRh$_{1-x}$Ir$_x$In$_5$ where $T_N>T_c$, \slrr\ below \tc\ initially drops as $T^3$, then crosses over to $T$ at some temperature $T<T_c$.  This might be understood as some portion of Fermi surface that remains ungapped.  In the \cecoincdx, the low temperature behavior behaves as $T^{1/4}$, but it is not clear why this should be so.  However, if the antiferromagnetism and superconductivity coexist microscopically, then it is clear that there are ordered local moments that create a field at the nuclear sites.   Goldstone fluctuations of the antiferromagnetic order can give rise to fluctuating hyperfine fields at the nuclear sites that potentially can relax the nuclei.   At present, it is not clear if the mechanism of the \slrrtext\ in this state arises from these magnons or from itinerant quasiparticles from gapless regions of the Fermi surface. Evidence for the former scenario can be found in comparisons of \slrr\ at the In(2) versus In(1) site in CeRh$_{1-x}$Ir$_x$In$_5$, which show an onset temperature for the $T$-linear behavior higher for the In(2) than for the In(1) \cite{ZhengCeRhIrIn5}. Since the static hyperfine field at the In(2) is higher, it is reasonable that the $T^{-1}_{1,magnon}$(In(2))$>T^{-1}_{1,magnon}$(In(1)), and therefore becomes the dominant relaxation mechanism at a higher temperature.  If the relaxation mechanism were purely from excited itinerant quasiparticles, then the temperature dependence of the In(1) and In(2) would be identical.

\subsection{Impurities in Heavy Electrons}
\subsubsection{Radioactive decay in \pucoga}

Superconductivity responds to impurities in a predictable fashion that depends on the nature of the superconductivity \cite{balatskyRMP,tinkham}.  Since NMR is sensitive to the density of states and is a microscopic probe, it is an ideal technique for investigating the nature of localized impurity states in superconductors and correlated electron systems \cite{AlloulHirschfeld}.  In a d-wave superconductor, pair breaking impurities locally suppress the superconducting order parameter and give rise to localized electronic states with energies below the superconducting gap.  With increasing impurity concentration, these localized states can overlap and develop into an impurity band that can contribute to the \slrrtext.  In many d-wave superconductors, this impurity band short-circuits the $T^3$ behavior and for sufficiently low temperature, \slrr\ crosses over to $T$-linear behavior \cite{imaiYBCOT1}.  A clear example of this behavior is seen in \pucoga, where the radioactive decay of the Pu strongly affects the superconductivity.  As $^{239}$Pu$\rightarrow \,\,^{235}$U + $^{4}$He has a half life of 24,110 years, significant lattice damage can build up in this material over time as the recoil of the heavy U nucleus, with a kinetic energy of 86keV, displaces several thousand atoms and the energetic alpha particle deposits its energy in the lattice \cite{wolfer}.  For sufficiently high temperatures, the lattice damage and Frenkel pairs can anneal out \cite{jutierPuCoGa5aging}.  Over time, however, pair breaking impurities suppress the superconductivity and \tc\ is reduced at the rate of $dT_c/dt \approx 0.2$ K/month.  Microscopically, these pair breaking impurities give rise to an impurity band near $E_F$ and prevent the density of states from reaching zero at $E_F$ as it would in a pristine superconductor.  This increase in $N(0)$ is reflected as an increase in \slrr\ and gives rise to a linear term that becomes important at sufficiently low temperature, and a finite $K$ as $T\rightarrow0$, rather than $K\rightarrow0$ as $T\rightarrow0$.  This effect is clearly observed in Knight shift and \slrr\ measurements in \pucoga; indeed the size of these effects correlates well with the age of the sample and the decrease of \tc\ with time \cite{CurroPuCoGa5}.

\subsubsection{Magnetic droplets in quantum critical systems}

\begin{figure}
\centering
\includegraphics[width=\textwidth]{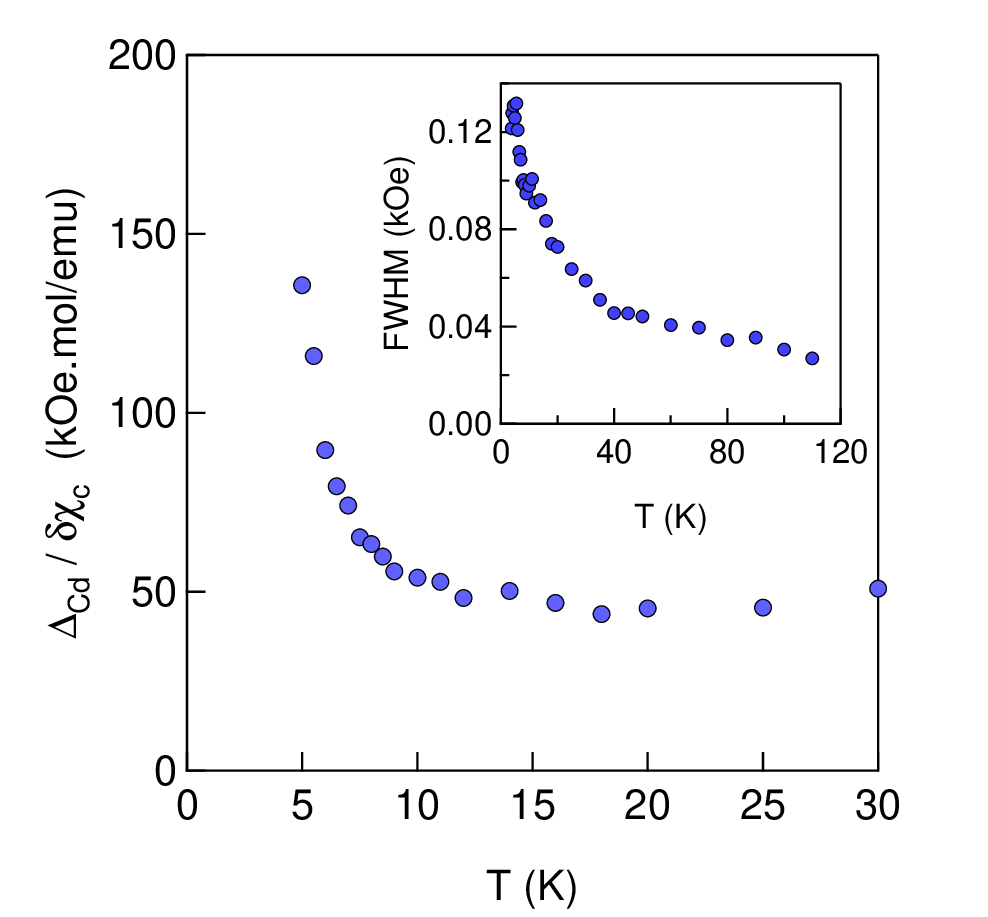}
\caption{A plot of $\Delta_{\rm Cd}(T)/M_{\rm Cd}(T) = \sqrt{\Delta^2(T) - \Delta_0^2}/(\chi_{\rm Cd}(T) - \chi_0(T)) \sim \xi_{SF}$ versus temperature in \cecoincdfifteen. The Inset shows the FWHM, $\Delta(T)$, of the Co resonance in the same material. Clearly, the antiferromagnetic correlations begin to grow in below $\sim$10 K. \label{fig:linewidth} }
\end{figure}

Clearly, NMR can shed important light on the nature of the superconductivity by probing how the local density of states is modified by the presence of impurities.  Another clear example in which NMR can elucidate the nature of the physics associated with impurities is in doped quantum critical systems \cite{CastroNetoNFL,SchmalianGriffiths,schmalianlocaldefect,castronetolocaldefect}. In a system that is close to a quantum phase transition, impurities may induce local perturbations that nucleate droplets of the ordered phase.  NMR in such a system may be sensitive to this local order.  In Cd doped \cecoin, the Cd nucleates antiferromagnetic order with radius $\xi_{AF}$.  For sufficiently high doping, these droplets interact with one another giving rise to long range antiferromagnetism \cite{FiskCddoping}.  The Cd doping also suppresses superconductivity, and above a critical concentration of dopants the system is purely antiferromagnetic.  The dynamics of this change in behavior is clearly visible in the \slrr\ of the In(1). Furthermore, since the Cd dopants nucleate antiferromagnetism, the local spin magnetisation develops a staggered response, similar to the Friedel oscillations induced by impurities in a Fermi liquid \cite{AlloulHirschfeld}.  The Co in this system is particularly sensitive to this staggered response, since it is located in a symmetric site between nearest neighbor Ce atoms.  The Co experiences a transferred hyperfine interaction to these two Ce spins, so that the form factor at the Co site vanishes for antiferromagnetic correlations.  The hyperfine field vanishes, however, only if the two Ce spins are equal in magnitude.  Around an impurity, the magnetization will decay within the correlation length (the details of how this order varies spatially depend critically on the nature of the quantum critical system) \cite{schmalianlocaldefect}.  Therefore, Co atoms sitting within such a droplet will experience finite hyperfine fields.  This distribution of fields is manifest as a broad linewidth that varies with doping and temperature (see inset Fig. \ref{fig:linewidth}). For non-interacting droplets, one can show that the linewidth, $\Delta$, is given by $\sqrt{\Delta_0^2 + \Delta_{\rm Cd}^2(T)}$ where $\Delta_{\rm Cd}(T) \sim M_{\rm Cd}(T)\xi_{AF}(T)$.  Here $\Delta_0$ is the intrinsic Co linewidth (Eq. \ref{eqn:dipolarsecondmoment}), and $M_{\rm Cd}(T)$ is the magnetization of the droplets.  Fig. \ref{fig:linewidth} shows $\Delta_{\rm Cd}(T)/M_{\rm Cd}(T)$ as a function of temperature, which clearly shows that $\xi_{AF}$ increases below $\sim 10$ K. A complete analysis is not available in this case, however, since the Cd concentration ($x=$1.5\%) is high enough that the droplets interact.  Similar effects were observed in the Ni and Zn doped cuprates \cite{bobroff,alloulreview,ouazi,morrhaaseanalysisYBCO}.  This new method promises to shed new light on the dynamics of the antiferromagnetic correlations that develop in these and other heavy electron materials.

\subsection{Summary}
At the time of this writing, a new class of iron oxypnictide superconductors has just been discovered, and NMR is playing a central role in experimental studies of the superconductivity and antiferromagnetism in these compounds.  Clearly, NMR has the potential and versatility to shed important light on the physics of these systems, as well as other classes of materials that may be discovered in the future.  There remain still several open questions in the study of the heavy fermions, though, that can be addressed with NMR.  In particular, the interplay between antiferromagnetism and superconductivity in different classes of materials and how these order parameters emerge for different classes of quantum critical behavior remains a fascinating area of research.  Secondly, the behavior of non-centrosymmetric superconductors is an active area of research for NMR \cite{MukudaNonCentro}.  And finally, the technique of purposefully adding impurities to these strongly correlated systems as a means to investigate the local spectroscopy of the impurity states with NMR holds tremendous promise to shed light on the nature of the order parameters that can emerge.